\documentclass[iop]{emulateapj}
\usepackage{color}
\usepackage{lscape}
\usepackage{subfigure}
\usepackage{natbib}
\usepackage{amsmath}
\slugcomment{{\sc Accepted to \apj:} December 13, 2010}
\shorttitle{UVX H$\alpha$}
\shortauthors{Adams \it{et al.}}

\begin{document}
\title{A new \lowercase{$z$}~=~0 metagalactic UV background limit\altaffilmark{*}}
\author{Joshua J. Adams}
\affil{Department of Astronomy, University of Texas at Austin, 1 University Station, C1400, Austin, TX 78712}
\email{jjadams@astro.as.utexas.edu}
\author{Juan M. Uson}
\affil{Observatoire de Paris - LERMA, 61 Avenue de l'Observatoire, 75014 Paris, France}
\author{Gary J. Hill and Phillip J. MacQueen}
\affil{McDonald Observatory, University of Texas at Austin, 1 University Station, C1402, Austin, TX 78712}
\altaffiltext{*}{This paper includes data taken at The McDonald Observatory of The University of Texas at Austin.}
\begin{abstract}
\par We present new integral-field spectroscopy in the 
outskirts of two nearby, edge-on, late-type galaxies 
to search for the 
H$\alpha$ emission that is expected from the exposure of their 
hydrogen gas to the metagalactic ultraviolet background (UVB).
Despite the sensitivity of the VIRUS-P spectrograph on the McDonald 2.7m telescope to low surface 
brightness emission and the large field-of-view, we do not 
detect H$\alpha$ to 5$\sigma$ upper limits of 6.4$\times10^{-19}$ erg/s/cm$^2$/\sq\arcsec\ in UGC~7321
and of 25$\times10^{-19}$ erg/s/cm$^2$/\sq\arcsec\ in UGC~1281 in each of the hundreds of 
independent spatial elements (fibers). 
We fit gas distribution models from overlapping 21 cm data of HI, extrapolate 
one scale length beyond the HI data, and estimate 
predicted H$\alpha$ surface brightness maps. We analyze three types of limits from the data 
with stacks formed from increasingly large spatial regions and compare to the model predictions: 
1) single fibers, 
2) convolution of the fiber grid with a Gaussian, circular kernel (10\arcsec\ full width 
half maximum), and 3) the coadded spectra from a few hundred fibers 
over the brightest model regions. None of these methods 
produce a significant detection ($>5\sigma$) with the most stringent constraints on the 
HI photoionization rate of
$\Gamma(z=0)<1.7\times 10^{-14}$ s$^{-1}$ in UGC~7321
and $\Gamma(z=0)<14\times 10^{-14}$ s$^{-1}$ in UGC~1281. 
The UGC~7321 limit is below previous measurement limits and also below current theoretical models. 
Restricting the analysis to the fibers bound by the HI data leads to a comparable limit; 
the limit is $\Gamma(z=0)<2.3\times 10^{-14}$ s$^{-1}$ in UGC~7321. 
We discuss how a low Lyman limit escape fraction in $z\sim0$ redshift star forming galaxies 
might explain this lower than predicted UVB strength and the prospects of deeper 
data to make a direct detection.
\end{abstract}


\section{Introduction}
\par The strength of the metagalactic ultraviolet background (UVB) has 
great impact on theoretical models of structure
formation \cite[e.g.][]{Haa96} and a variety of physical processes such as
the inhibition of small halo collapse \cite[e.g.][]{Efs92}, the intergalactic
temperature and ionization state of the intergalactic medium (IGM) \cite[e.g.][]{Hui97}, and
IGM metallicity determinations \cite[e.g.][]{Rau97a}. The likely contributors
to the UVB are active galactic nuclei and star formation in galaxies 
\citep{Sch03,Fau08a} with appear compatible with observed populations
\citep{Gal95,Hop04,Hop07,Bou09} under reasonable corrections for dust 
attenuation, low luminosity extrapolations, redshift evolution, and
escape fractions. The strength of the UVB, especially at low redshift \citep{Dav01}, 
is still highly uncertain despite its importance. Most 
recent efforts have focused on high redshifts, $z>2$, where 
the strongest UVB measurements exist. For instance, the detailed history 
of star formation \citep{Mad99,Fau08b} and the potential to measure 
individual active galactic nuclei (AGN) host halo masses 
\citep{Loe95,Fau08c} have been explored. 
Measurements of the photoionization rate have used
three methods: observations of H$\alpha$ such as 
described in this paper, the line-of-sight proximity effect 
method \cite[e.g.][]{Car82,Bat88}, and the flux decrement method \cite[e.g.][]{Cen94,Rau97b}. The 
latter two require backlighting quasars and are therefore 
difficult or impossible at low redshift. 
We are motivated to constrain the current model with a different, low redshift 
measurement. Instead of using Lyman-$\alpha$ forest features, we 
pursue a measurement of the UVB powered, H$\alpha$ 
emission that should occur in the outskirts of local disk galaxies. 
As a secondary motivation, the kinematics of H$\alpha$ at distances 
beyond HI data are important probes to the total dark halo masses in 
nearby disk galaxies \citep{Chr08}. 
\par Galactic disks are optically thick to Lyman 
limit photons and maintain their observed HI distributions 
through self-shielding against the UVB. As 
recognized for decades \citep{Sun69,Fel69,Boc77}, 
the influence of the UVB may be 
investigated in the extreme outskirts of disks where the 
self-shielding begins to fail. These early works sought to 
measure this effect through disk truncation in HI. However, 
there appear to be cases with \citep{Cor89,vGo93} and without 
\citep{Wal97,Car98,Oos07} HI truncations 
above the critical column density predicted using current 
UVB estimates, implying that other processes may 
strip gas and mimic the result. Moreover, reaching the 
UVB implied truncation thresholds in 21 cm measured HI would require 
rather long observations with current facilities. 
A more robust signature 
of the UVB strength would be the detection of the H$\alpha$ in 
these outskirt regions. H$\alpha$ has been found at 
such radii before in actively star forming and warped 
galaxies by \citet{Bla97} (hereafter BFQ) with Fabry-Perot staring measurements. 
However, the $\mu(H\alpha)=2.3\times$10$^{-19}$ erg/s/cm$^2$/\sq\arcsec\ 
detection was 
interpreted to be due to non-UVB sources as indicated by an abnormally high 
[NII]$\lambda$6548 to H$\alpha$ ratio. Searches have also yielded limits 
in quiescent systems \citep{Vog95,Wey01,Mad01} 
with an upper limit for the UVB photoionization rate, $\Gamma$, of 
$\Gamma(z=0)<2.4-9.5\times10^{-14}$s$^{-1}$(2$\sigma$) 
being the deepest. The wide range due on this limit is due to gas cloud geometrical uncertainty. 
Despite the numerous theoretical implications and the efforts of numerous groups, a 
UVB powered H$\alpha$ detection still awaits discovery. 
\par The tactical advantages we bring to this problem are deep surface brightness 
limits, a large two dimensional field of view through integral field spectroscopy compared 
to the previous longslit and Fabry-Perot staring data, and target selection of 
very high inclinations to 
maximize signal and minimize contamination uncertainty. Our targets are edge-on, low 
surface brightness Sd galaxies that are rather
isolated and minimally warped in order to avoid density distribution uncertainties and
exposure to internally generated ionization from smaller radii. 
Indeeed, our most constraining target, UGC~7321,
has a gas surface density below that required for significant star formation
\citep{Ken89} at all radii, as well as being unusually isolated with
no known companions and minimal ($<3\arcdeg$) warping \citep{Uso03}. 
\par In this paper we begin with a description of the 
simple ionization state and density model of disk galaxies 
that will be used to link a measured H$\alpha$ surface 
brightness with a particular UVB photoiozation rate in 
\S \ref{sec_mod_HI}. 
In \S \ref{sec_mod_fit}, we give disk parameter constraints 
based on fits to existing 21 cm data. 
In \S \ref{sec_alt_lim}, we argue that UGC~7321 in 
particular is likely to extend its HI profile beyond the 
current 21 cm limits without truncation. 
In addition, the HI observations of UGC~7321 are amongst the most 
sensitive such measurements published to-date. The 21 cm data allow 
a very precise model to be made for the gas distribution in the 
galaxy outskirts at the locations where we search for 
H$\alpha$ emission. 
Next, in 
\S \ref{sec_data}, we present deep integral field spectroscopy 
observations at radii corresponding to the outermost 
detections of 21 cm emission and beyond. 
We describe the choices made to stack spectra on various spatial scales. 
The stacked spectra are searched for H$\alpha$ detections and upper limits are derived. Particular focus is 
given to systematic errors. Finally, in \S \ref{sec_dis}, we discuss the context, the likely cause 
of the unexpectedly low limit, and further observations that can 
confirm our conclusions. The Appendix \ref{sec_full_mu} provides the analytic 
details necessary to construct the full and general H$\alpha$ 
surface brightness distribution model. We will quote most of the surface 
brightness limits in units of erg s$^{-1}$ cm$^{-2}$ arcsec$^{-2}$, but 
for easy comparison to alternative units we note the 
conversion at the wavelength of H$\alpha$ of 
1 millirayleigh (mR)$=5.66\times 10^{-21}$ erg s$^{-1}$ cm$^{-2}$ 
arcsec$^{-2}=2.8\times 10^{-3}$ cm$^{-6}$ pc in 
emission measure assuming the case B coefficient we adopt. 
\section{HI based models and H$\alpha$ predictions}
\label{sec_mod}
\subsection{Model assumptions}
\label{sec_mod_HI}
\par A three dimensional gas density distribution must be inferred 
in order to translate H$\alpha$ surface 
brightness into a UVB strength. BFQ made estimates 
assuming exponential forms both radially and 
vertically in the 
gas distribution with a plane parallel assumption. Motivated by the 
regular HI structure on local scales \citep{Gar02,Uso03} of our 
chosen targets showing simple exponential trends and needing an 
extrapolated model in gas density 
for interpretation of UVB limits, we also assume exponential forms. 
\par In order to interpret H$\alpha$ measurements generically 
inside and outside of the UVB photoionization front around 
gaseous disks, we have generalized the model of BFQ. Some toy 
calculations in the model also show the importance of 
high inclination selection to make the deepest possible 
UVB constraints. This high inclination boon has been known 
before, but not carefully followed in earlier works' target selection. 
The model assumes both regular gas distributions and sharp 
photoionization transitions in a plane parallel approximation under 
arbitrary disk inclinations and sight lines. Our model 
assumes sharp photoionization fronts exist. We verify this 
assumption by estimating the Lyman limit photon mean free path 
at the midplane ionization front. In their Equation 3, BFQ 
estimate the hydrogen density at this point as $n_{H}\approx0.05$cm$^{-3}$. 
The Lyman limit photon mean free path is given by 
$l_{mpf} \approx (n\times a_{\nu})^{-1} \approx 1.1$pc with $a_{\nu}$
\citep{Ost06} as the
hydrogen Lyman limit photoionization cross section. 
This is much smaller than the common disk 
scale lengths in either direction. The vertical 
scales for cold disk galaxies are of order 100 pc or greater. 
More sophisticated models can be made \citep{Mal93,Dov94} by 
solving for the ionization and excitation states of 
hydrogen and helium with full radiative transfer solutions in a 
grid of plane-parallel gas layers, but such an analysis is beyond the 
scope of this work. 
\par The forthcoming derivation follows BFQ equations~1-6. The important 
differences are that this derivation is generalized for any 
viewing inclination, $i$, and for arbitrary positioning of the 
spectral data in the galaxy's field of observation. 
The BFQ derivations were specifically for $i=0$\arcdeg\ and 
the field position along the major axis where all gas is photoionized. 
We denote the generic surface brightness in H$\alpha$ as $\mu$. 
We denote $\mu_0$ as the special case of the peak H$\alpha$ surface brightness 
where the photoionization front intersects the disk midplane. Our results reduce to the 
BFQ values of $\mu_0$ for $i=0$\arcdeg. In Equation \ref{eq_dens} we give the 
assumed gas distribution in 
cylindrical coordinates R and z with radial scale-length 
h$_r$, vertical scale length h$_z$, and central hydrogen density $n_0$. 
\begin{equation}
\label{eq_dens}
n_H(R,z)=n_0 \exp(\frac{-|z|}{h_z}) \exp(\frac{-R}{h_r})
\end{equation}
The commonly assumed form of the UVB spectrum is given in 
Equation \ref{eq_spec} where $\nu$ is the frequency, 
$\nu_0$ is the Lyman limit frequency, $J_{0}$ 
is the UVB strength at the Lyman limit in units of 
erg cm$^{-2}$ s$^{-1}$ Hz$^{-1}$ sr$^{-1}$, 
and $\beta$ is the UVB spectral index. 
\begin{equation}
\label{eq_spec}
J_{\nu}=J_{0}\left( \frac{\nu_0}{\nu} \right) ^{\beta}
\end{equation}
Another common form of quoting the UVB strength is with the 
UVB photoionization rate, $\Gamma$. We show this form in Equation \ref{eq_phot} 
where $h$ is Planck's constant, $\sigma(\nu)$ is the 
hydrogen photoionization cross section, and $a_{\nu}$=$\sigma(\nu_0)$ is the 
Lyman limit cross section. The final equality in Equation 
\ref{eq_phot} comes from the standard power law 
approximation to the cross section shape \citep{Ost06}. 
\begin{equation}
\label{eq_phot}
\Gamma=4\pi \int^{\infty}_{\nu_{0}}\frac{J_{\nu}\sigma(\nu)}{h\nu} d\nu=\frac{4\pi a_{\nu} J_0}{h\times(3+\beta)}
\end{equation}
In Equation \ref{eq_ph_ion} we equate recombination and ionization rates under a 
plane parallel approximation. For the radial regions where
any self-shielding can take place, we consider the top and bottom
of the disk to each see incident flux from only half their
total solid angle. We 
define $n_e$ as the electron density, $n_p$ as the proton density, 
$\xi$ as the ionization fraction,
$\alpha_B$ as the case B recombination coefficient, and $z_c(R)$ as the
height above the midplane to which the photoionization front penetrates at radius R. 
We define $\varepsilon$ as the volume filling factor, assumed to be
spatially invariant. A clumpy gas distribution can, to first order, be
represented by using this term somewhat lower than the nominal value of unity.
With the assumption of sharp ionization boundaries, we can equate 
the gas densities as $n_e=n_p=\xi n_H$ at radii beyond the 
photoionization front. 

\begin{equation}
\label{eq_ph_ion}
\alpha_B\int^{\infty}_{z_c(R)} \xi^2 \varepsilon n_H^2(R,z) dz = \int^{\infty}_{\nu_{0}} \frac{2\pi J_{\nu}}{h\nu} d\nu = \frac{2\pi J_0}{h\beta}
\end{equation}
We next define a threshold radius, $r_c$, to which the UVB penetrates fully through the disk plane, so $z_c(r_c)=0$. The solution 
of Equation \ref{eq_ph_ion} leads to Equations \ref{eq_rc} and \ref{eq_zc}. 
\begin{equation}
\label{eq_rc}
r_c=(\ln(2 \xi^2 \varepsilon \alpha_B a_{\nu} n_0^2 h_z \beta)-\ln(\Gamma\times(3+\beta)))\times h_r/2
\end{equation}
\begin{equation}
\label{eq_zc}
z_c(R)=\left\{
\begin{array}{lr}
\pm(r_c-R)\times h_z/h_r & : R\leq r_c\\
0: R>r_c
\end{array}
\right.
\end{equation}
Next, we define the 
variable $\rho$ as the distance from the disk's midplane along the line-of-sight, 
spanning $-\infty$ to the observer and $\infty$ away from the observer. We 
also define the major axis position $b_1$, and minor axis position 
$b_2$ as the observed field positions projected onto the sky. Finally, 
we represent the galaxy's inclination with $i$. 
Simple transformations to 
cylindrical coordinates give the expressions 
in Equations \ref{eq_z_tran} and \ref{eq_r_tran}. 
\begin{equation}
\label{eq_z_tran}
|z|=\sqrt{\rho^2 \cos^2 i + b_2^2 \sin^2 i + 2\rho b_2 \sin i \cos i}
\end{equation}
\begin{equation}
\label{eq_r_tran}
R=\sqrt{\rho^2 \sin^2 i + b_2^2 \cos^2 i + b_1^2 - 2\rho b_2 \sin i \cos i}
\end{equation}
The H$\alpha$ surface brightness, $\mu(b_1,b_2)$, follows directly from a line-of-sight integration. 
The full evaluation of $\mu(b_1,b_2)$ involves finding the values of $\rho$ that 
intersect the photoionization surface described by $z_c(R)$ with field positions $b_1$ and $b_2$. 
The analytic solutions to those intersections are given in 
Appendix \ref{sec_full_mu}. The solution for the special case at field position $b_2=0$ and 
$b_1=r_c$ gives the aforementioned peak surface brightness $\mu_0$, which is itself 
a useful measurement parameter, as related in Equation \ref{eq_SB}. In Equation \ref{eq_SB}, 
$\gamma$ is the generally non-analytic integration of the emissivity 
along the line of sight, $\alpha^{\mbox{eff}}_{H\alpha}$ is
the case B effective H$\alpha$ recombination coefficient, and $\Omega$ is the full sky solid angle of
4$\pi$ sr. For the gas density parameters we derive in our target galaxies in 
\S \ref{sec_mod_fit} and the areas we observe in \S \ref{sec_data}, the 
face-on column densities of total hydrogen are sufficient ($>10^{17}$ cm$^{-2}$) to be 
everywhere optically thick to Lyman limit photons, let alone Lyman-$\beta$ 
and the other important lower energy transitions. 
We use $\alpha^{\mbox{eff}}_{H\alpha}=1.17\times10^{-13}$ cm$^{3}$ s$^{-1}$ as 
appropriate for T$=10^4$K \citep{Ost06}. 
\begin{eqnarray}
\label{eq_SB}
\mu_0=\frac{2 h \nu_{H\alpha} \alpha^{\mbox{eff}}_{H\alpha} \int^{\infty}_{0} n_e(R,z) n_p(R,z) d\rho}{\Omega}= \nonumber \\
\frac{2 \xi^2 \varepsilon \alpha_{H\alpha}^{\mbox{eff}}n_0^2 h \nu_{H\alpha} \gamma}{\Omega} \nonumber \\
\mbox{with } \gamma= \nonumber \\
\int^{\infty}_{0}\exp(-\frac{2\rho \cos i}{h_z}-\frac{2\sqrt{r_c^2+\rho^2 \sin^2 i}}{h_r}) d\rho
\end{eqnarray}
\par We explain the use of certain constants and assumed values to Equation \ref{eq_SB}. 
The ionization fraction is assumed to be unity by the earlier discussion 
of the Lyman limit mean free path. The volume filling factor may approach 
unity as there is no indication of star formation at extended 
scales in these galaxies. We will discuss the evidence for the absence of 
extended star formation in Section \ref{sec_dis}. Furthermore, the deprojection 
of the HI distribution in \citet{Uso03} gives a peak surface density of 
only 5.8 $M_\odot$ pc$^{-2}$ at the center of UGC~7321. The surface density drops by over 
an order of magnitude at the locations we observe. These surface densities are well below the dynamical 
criterion for efficient star formation \citep{Ken89} and make a 
smooth gas distribution plausible. It is not possible to exclude small scale clumpiness, so we retain the 
volume filling factor. The case B and H$\alpha$ effective recombination 
coefficients are dependent on electron temperature. Following \citet{Wey01} and 
the discussion therein, we adopt T=10,000 K and the values of \citet{Ost06}. 
The true electron temperature may plausibly be different by a factor of two, leading to 
corresponding changes in $\alpha_{H\alpha}^{\mbox{eff}}$ and 
$\alpha_{B}$ of the same order of magnitude. However, the linearization 
in $\Gamma$ of Equation \ref{eq_SB} 
makes the surface brightness depend on the ratio of these two recombination 
coefficients, so their similar behavior with electron temperature cancels. For consistency 
with previous works, we do not propagate the recombination coefficient uncertainties 
as systematics to the final UVB limit. 
\par
\begin{figure}
\centering
\includegraphics [scale=0.8,angle=-90]{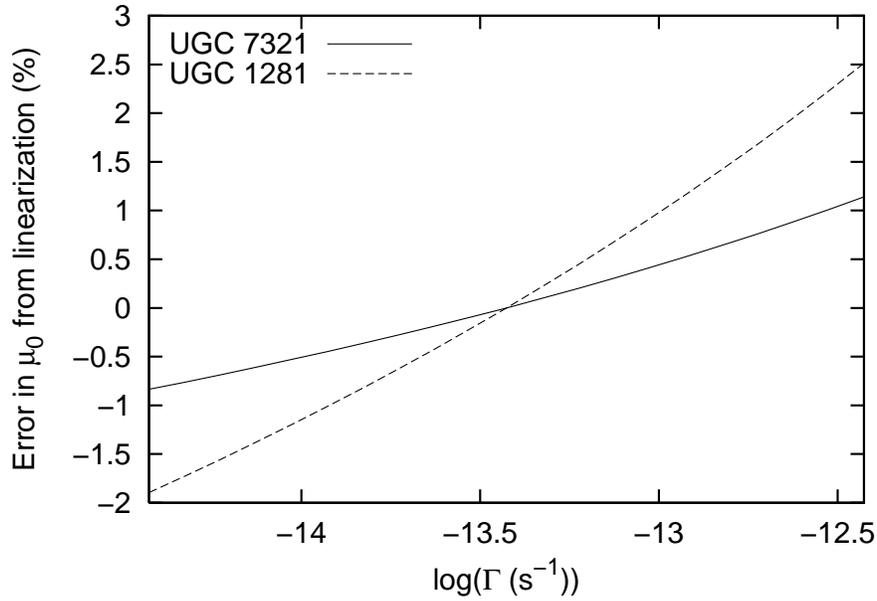}
\caption{The fractional error in a linear relation between the H$\alpha$ peak 
surface brightness and the UVB photoionization rate, $\Gamma$, 
under different UVB strengths. The 
parameters for the two target galaxies are given 
in Table \ref{tab_MC} and their derivation described in 
\S \ref{sec_mod_fit}. The pivot in the linearization at 
$\Gamma=4\times10^{-14}$ s$^{-1}$ represents the 
current best estimate from \citet{Fau09}. 
}
\label{fig_lintest}
\end{figure}

 Some brief numerical examples set the expected orders of magnitude, 
quantify the achievable limits under different galaxy geometries, and 
illustrate the important parameter dependencies under linear expansions. 
We look at some trial cases with
$\varepsilon=1$, $h_r=1000$ pc, $h_z=100$ pc, $\beta=1.8$,
$\Gamma=4\times 10^{-14}$ s$^{-1}$, and $n_0=5$ cm$^{-3}$. For $i=0\arcdeg$, 
$\gamma=\exp(-2 r_c/h_r)\times h_z/2$ so $\mu \approx 3.0\times 10^{-20}$ 
erg/s/cm$^2$/\sq\arcsec. 
For $i=90\arcdeg$, $\gamma=r_c K_1(2 r_c/h_r)$. $K_1(x)$ is the modified Bessel function. 
In this case, $\mu \approx 1.3\times 10^{-18}$ erg/s/cm$^2$/\sq\arcsec. 
For this work's applications, the surface brightness profiles are smoothed by
seeing and sampled by large fibers. Realistic smoothing and sampling,
of order several arseconds, can lower these peak values by several tens of
percent. 
We will assume for all 
calculations that $\beta=1.8$ as 
motivated by previous models \citep{Shu99} and to aid the comparison with 
previous observational work that used the same assumption \citep{Wey01}. 
We note that $\mu_0$ 
scales exactly linearly with $\Gamma$ when viewed face-on and 
nearly linearly for all other inclinations. This is easily demonstrated by 
taking the large argument asymptotic behavior of the modified Bessel function which 
yields a linear scaling in $\Gamma$ after a first order expansion. We show 
the small error caused by assuming a linear relation between 
$\mu_0$ and $\Gamma$ in Figure \ref{fig_lintest} for reasonable 
geometries. All further 
estimations of $\Gamma$ in this work will be made in the linear approximation. 
We have linearized our estimate around $\Gamma=4\times 10^{-14}$ s$^{-1}$ 
because we consider it the best current estimation from the work of 
\citet{Fau09}. However, any initial value would have worked as the only effect
of a particular choice is that the small non-linearities pivot around
the simulation UVB choice, but this error is negligible compared to our other error 
terms. 
The discussed numerical example between $i=0\arcdeg$ and $i=90\arcdeg$ 
also shows how the selection of thin, edge-on disks can exploit a particular flux 
limit to a ($30-50\times$) stronger UVB constraint than for face-on disks. 
We also emphasize with Equation \ref{eq_SB} that the first order effects 
near $i$=90\arcdeg\ on distance, volume filling factor, and gas density cancel out; 
$\mu_0$ only has first order dependence on $i$, the ratio of scale lengths, $\frac{3+\beta}{\beta}$, and 
$\Gamma$. 

\subsection{HI data}
\label{sec_mod_fit}
\par Three-dimensional gas distributions must be inferred for individual galaxies to 
interpret H$\alpha$ surface brightness and to guide the stacking choices 
amongst fibers. We will use such fits to extrapolate the density 
profiles to larger radii where the H$\alpha$ emission is 
predicted to reach peak surface brightness. The parameters from 
stellar distributions could potentially be used, but 21 cm measured HI 
is the more relevant indicator to ionized hydrogen. We adopt distances of 
10 Mpc for UGC~7321 \citep{Uso03} and 5 Mpc for UGC~1281 \citep{Tul06}. 
Low redshift surface brightness is insensitive to distance, so the 
exact distances are unimportant to this work. Different literature 
estimations have 50\% and 10\% rms ranges for 
the UGC~7321 and UGC~1281 distances respectively. 
We indicate scale lengths by the terms 
$d_{10}$ as the actual distance to UGC~7321 in units of 10 Mpc and 
$d_5$ as the actual distance to UGC~1281 in units of 5 Mpc. For reference, 
the scale conversions become 48.5$d_{10}$ pc/\arcsec\ for UGC~7321 and 
24.2$d_{5}$ pc/\arcsec\ for UGC~1281.
\par UGC~7321 was observed by one of us in collaboration with L.~D.~Matthews
\citep{Uso03} using the second most-compact (C) configuration of the
VLA\footnote{The Very Large Array of the National Radio Astronomy
Observatory is a facility of the National Science Foundation, operated under cooperative
agreement by Associated Universities, Inc.} which
includes some of the shortest spacings available and their full coverage,
deep observations yielded spacings down to 28m, close to the dish diameter. 
Their quasi-naturally weighted (``robust'' parameter ${\cal R} = +1$) images recovered
the full single-dish flux and, moreover, their single-dish equivalent spectrum
matched the features of the best single-dish spectrum to within the (higher)
uncertainty of the single-dish observations (\citet{Uso03}, fig. 6). For this paper,
we have used their quasi-uniformly (${\cal R} = -1$) weighted images because of
their better resolution ($\sim$12\arcsec\ ) although the somewhat higher (45\%)
noise level only recovers $\sim$96\% of the total flux. However, the five 
parameter model fits to the zeroth moment maps, described below, recover some of the
lost flux and the remaining uncertainties only slightly shift the position
of the predicted H$\alpha$ peak.

\par For UGC~1281, we have reduced the raw data from the VLA archive.
It was observed under proposal AZ097 on 1997 December 26 in the most
compact (D) configuration for a total of 3 hours on source with interspersed
observations of the strong, primary calibrator J0137+3309 (3C48) for which
we
have adopted the VLA recommended flux density of 15.9~Jy. 
The observations were spaced over a range of $\pm 3$~hours in H.A. giving
excellent uv-coverage and images with 127 channels of width $\sim 2.6$ {\mbox{km s
$^{-1}$}} after standard on-line Hanning-smoothing. The angular scale that
corresponds to the shortest baseline ($\sim$900\arcsec) is sufficiently larger than the
largest single-channel galaxy extent ($\sim$285\arcsec) that the array should have
recovered the total HI flux. We followed the same reduction
steps as for UGC~7321 \citep{Uso03} to obtain a ``cube'' of spectral images
using
nearly-natural weighting (${\cal R} = +1$) which gave images with resolution
$\sim$51\arcsec\  which were free of artifacts to the rms sensitivity
$\sigma \sim$1.0~mJy/beam per channel. We computed moment maps after
applying a standard ``1-$\sigma$ cutoff'' evaluated on a cube Gaussian-
smoothed spatially to 70\arcsec\ and Hanning-smoothed in frequency which led to a
total HI flux of $41 \pm 2$~Jy {\mbox{km s$^{-1}$}} corresponding to a mass of
$2.3 \times 10^8$~$d_{5}^2 M_\odot$. The total flux is in good agreement
with the values in the literature which range from (35.8 to 38.9) Jy km/s from two
different
single-dish measurements \citep{Huc89} with the spread and uncertainty due in 
part to some ringing from strong in-band
HI emission from the Milky Way as well as to calibration uncertainties.
Again, we have obtained a spectral ``cube'' with nearly-uniform weighting
(${\cal R} = -1$) which gave images with resolution of $\sim$42\arcsec\ with
rms sensitivity $\sigma \sim$1.5~mJy/beam per channel.  As in UGC~7321, the
higher noise level results in a slightly lower total flux, $39 \pm 2$~Jy
{\mbox{km s$^{-1}$}}.

\par
\begin{figure}
\centering
\subfigure{\includegraphics [scale=0.4,angle=-90]{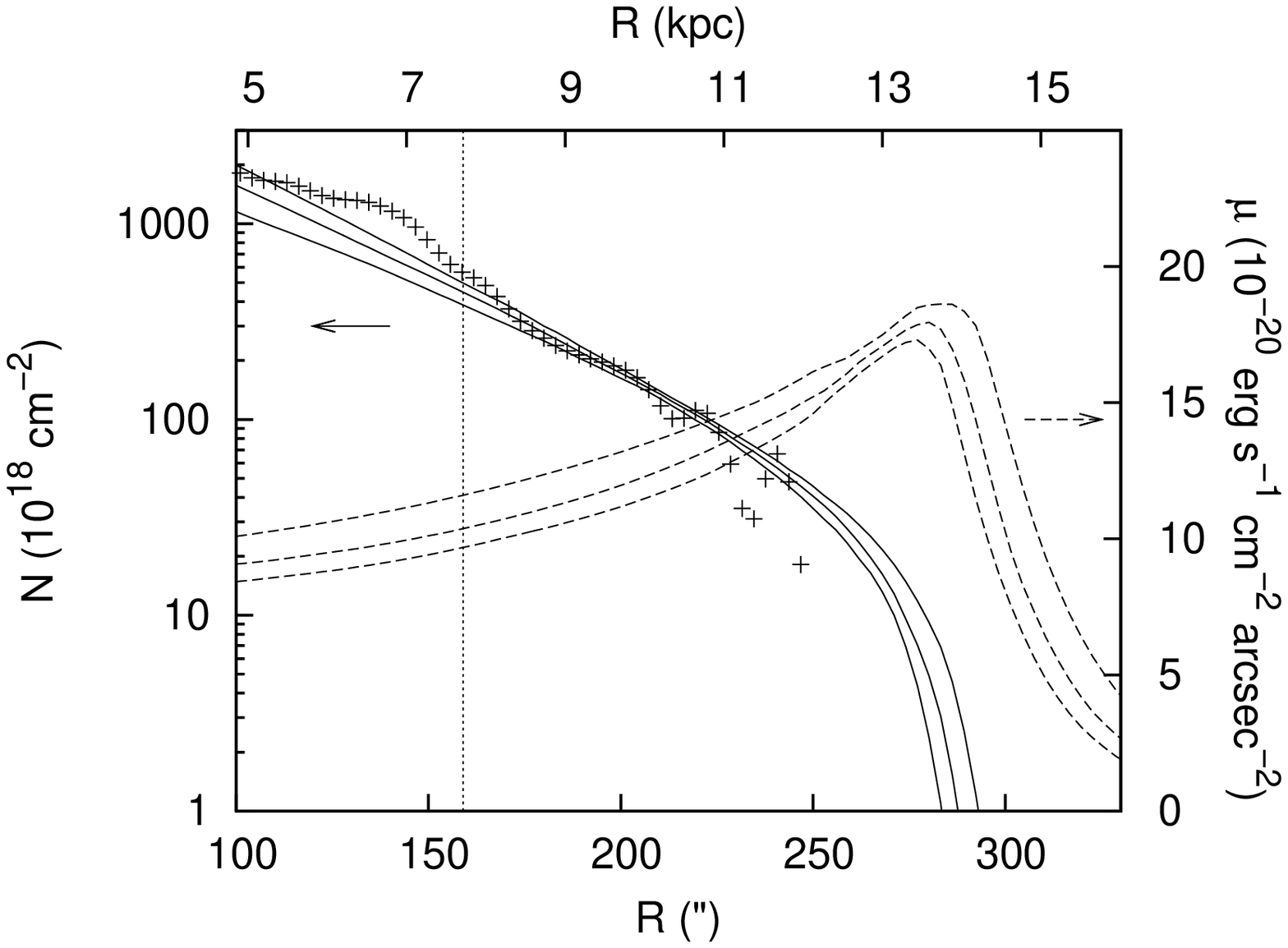}}
\subfigure{\includegraphics [scale=0.4,angle=-90]{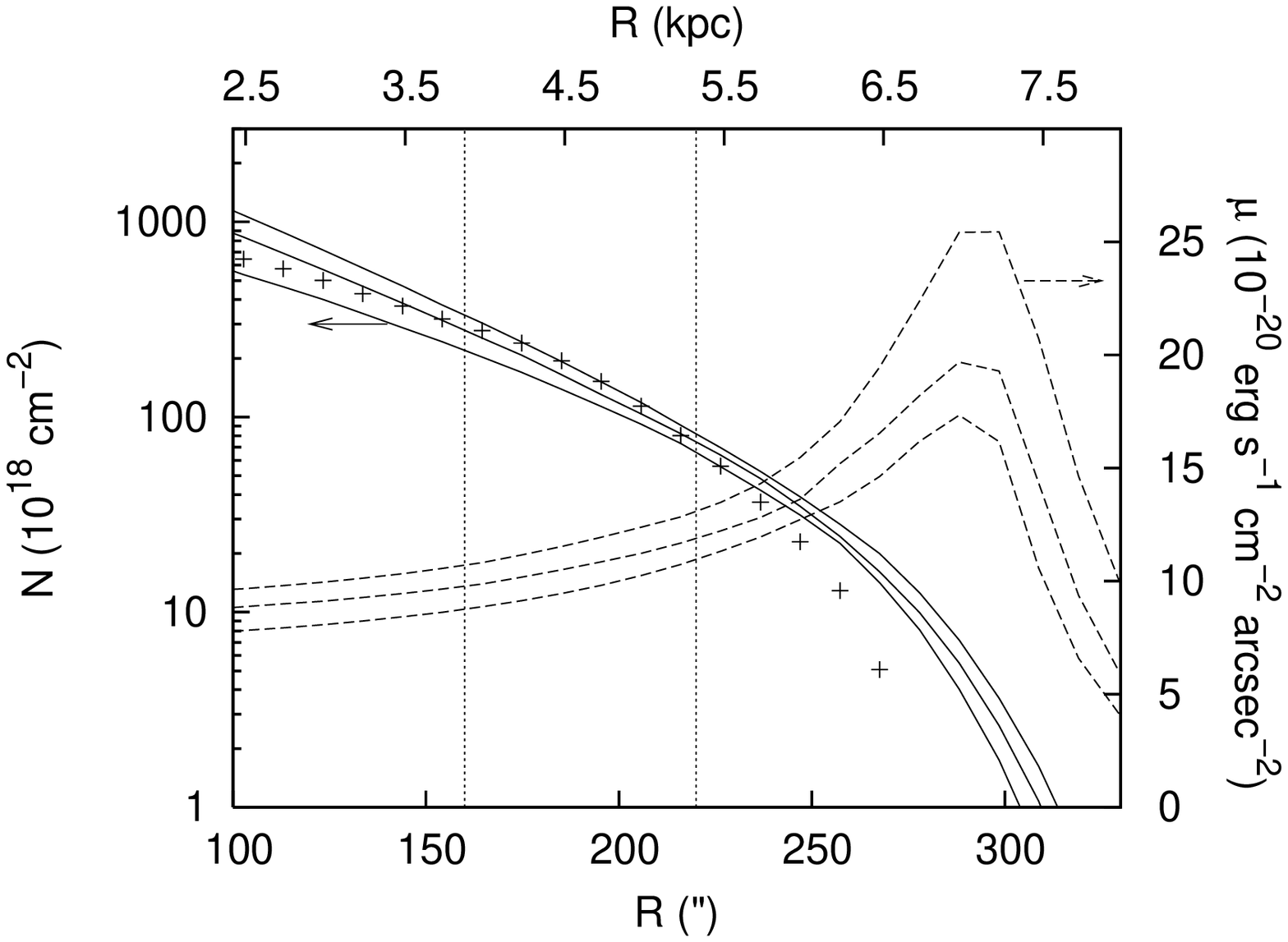}}
\\
\subfigure{\includegraphics [scale=0.4,angle=-90]{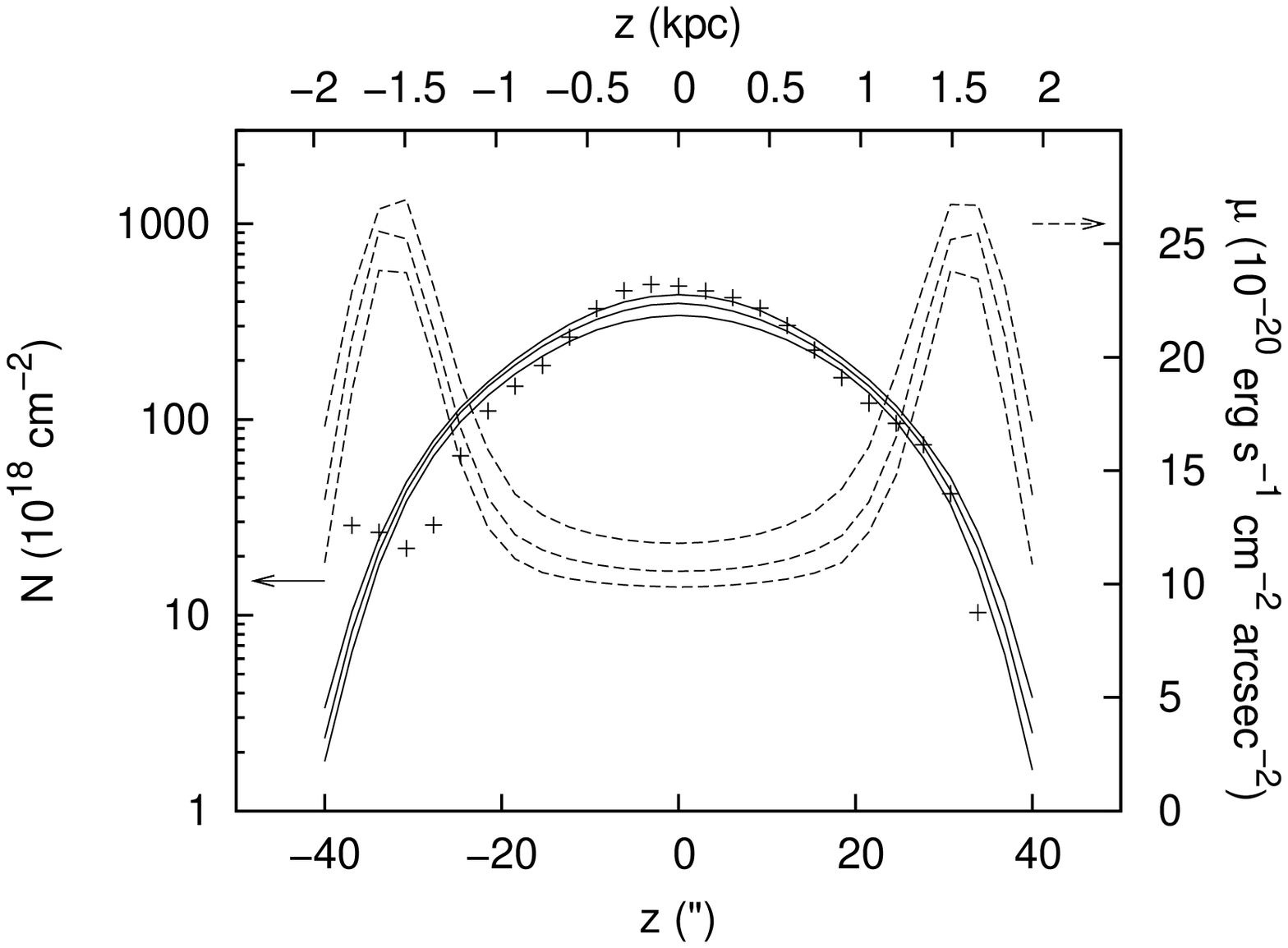}}
\subfigure{\includegraphics [scale=0.4,angle=-90]{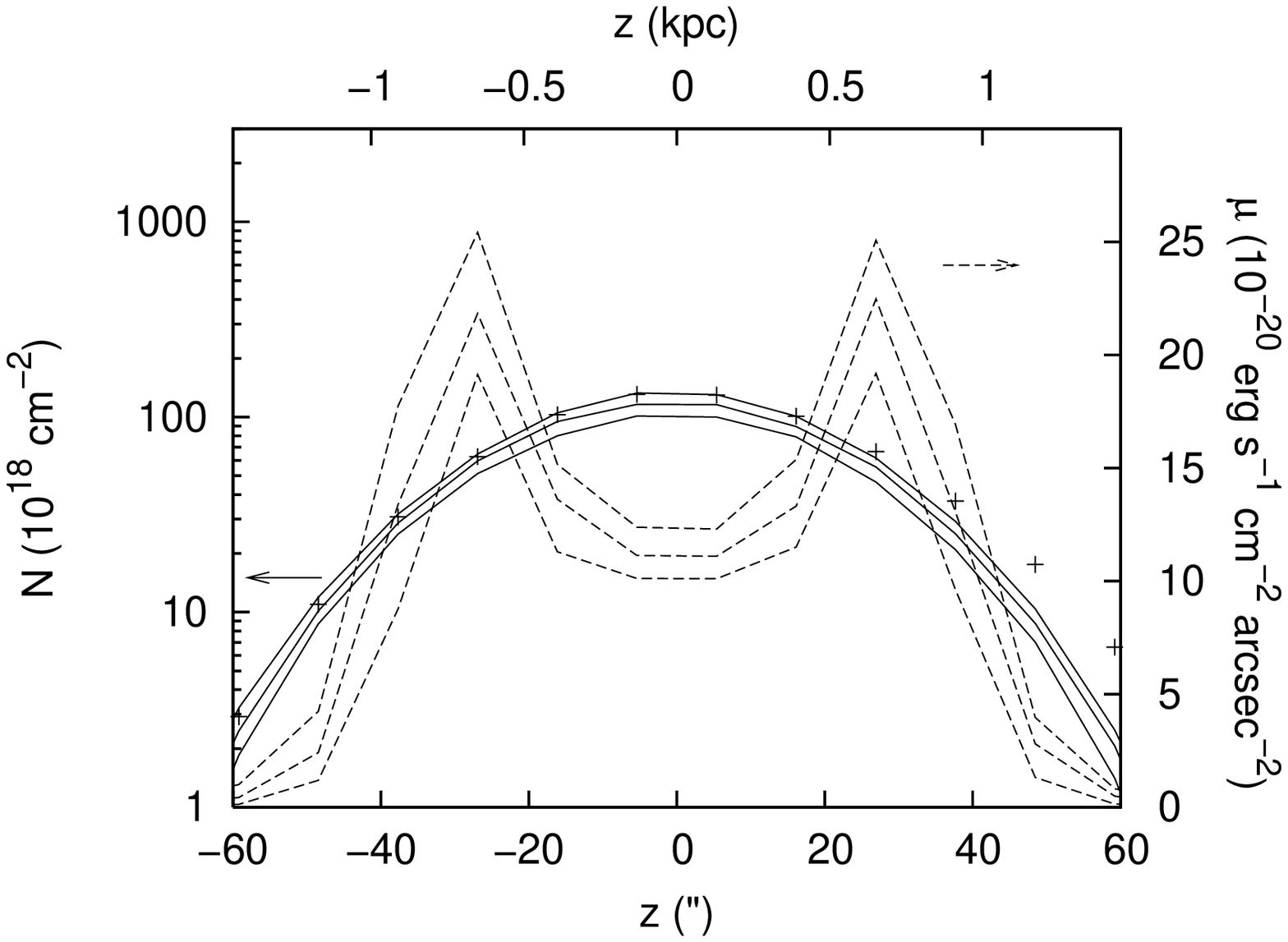}}
\caption{The single position angle, parameterized fits to the HI 
distributions shown as solid lines. Because we assumed a single position angle and single 
radial scale length, the ranges over which we fit the HI distributions 
must be somewhat controlled and limited to the radii near the 
H$\alpha$ observations. The predicted H$\alpha$ surface brightness 
profiles are also shown against the right side axis as dotted lines. The 
horizontal arrows indicate the correct axis for each profile. 
The breaks at large radii in the 21~cm profiles are due to intersections 
with the UVB photoionization fronts.
\textit{Top} The data along the major axes. 
\textit{Bottom} The data along the minor axes at particular 
offsets. 
\textit{Left} Cuts along the midplane and normal to it offset by
165\arcsec\ with data, models, and 68\% confidence intervals in UGC~7321. 
We restrict the fit to points $>160\arcsec$ from 
the galaxy's center as indicated by the vertical dotted line. 
Using the best parameters from Table \ref{tab_MC}, the 
threshold radius (Equation \ref{eq_rc}) with the nominal 
value of $\Gamma=4\times 10^{-14}$ s$^{-1}$ is predicted to be at 
$r_c=13.4$kpc. Our spectroscopic data cover regions from R$=9.5$kpc 
to R$=14.6$kpc. 
\textit{Right} Similarly, data and fits to 
UGC~1281. The offset here is 200\arcsec\ 
from the minor axis. The points between the dotted lines at 160\arcsec\ and 
220\arcsec\ form the restricted range of the fit as a 
$\approx 8\arcdeg$ warp becomes important beyond. This fit appears 
poorer because of the larger warp, but a fit to all points 
at R~$>$~220\arcsec\ returns the same radial scale length to within the Monte Carlo errors. 
Using the best parameters from Table \ref{tab_MC}, the
threshold radius (Equation \ref{eq_rc}) is predicted to be at
$r_c=7.4$kpc. Our spectroscopic data cover regions from R$=5.7$kpc
to R$=9.1$kpc. 
Since we only show one dimensional cuts, 
these figures do not show all the datapoints used in the fits.
}
\label{fig_MC}
\end{figure}

 Next, we characterize the HI distributions nearest our H$\alpha$ observations. 
We have derived five parameter fits in $n_{0}, h_{r}, h_{z}, i$, and 
position angle to the zeroth moment maps of UGC~7321 and 
UGC~1281 through non-linear least squares minimization. 
The models include convolution to the instrumental beams of 
$\sim$12\arcsec\ and $\sim$42\arcsec\ 
FWHM and sampling of H$\alpha$ appropriate to the fiber data. 
Both the maps show at least two major axis power law slopes, 
as \citet{Chr10} have found to be common in extended gaseous disk gas. 
We do not try to model the full gas distributions, but only the 
large radius trends by restricting the fits heavily to the 
outermost data regions.  Still, the model fits deviate from the data by an amount 
that exceeds the observational errors. Some minor warps and substructure are visible. 
The formal errors in the total line intensity images are
$15\times10^{18}$ cm$^{-2}$ for UGC~7321 and $5\times10^{18}$ cm$^{-2}$ for
UGC~1281, which are both far smaller than the residuals to the best fit models.  
In order to capture the systematic model errors, we have made Monte Carlo simulations 
between the data and the best fit models to create 
68\%\ confidence intervals as given in Table \ref{tab_MC} for all 
disk parameters and H$\alpha$ observables. 
The perturbations in the Monte Carlo simulations are made from the 
residuals of the best fit model, not the statistical errors, 
to include the influence of systematics. These simulations 
allow us to create three types of H$\alpha$ surface 
brightness prediction, with different scales of spatial co-addition, 
under an assumed $\Gamma$. 
Note that many of the individual disk parameters in Table 
\ref{tab_MC} have large relative uncertainties, but the 
surface brightness predictions have small relative 
uncertainties. The disk parameters share 
degeneracies, as captured in the Monte Carlo 
simulations, that create highly certain H$\alpha$ predictions \
despite the individually uncertain gas parameters. 
Predictions can be made for individual fibers, but to 
both mitigate the model uncertainties and 
improve our limits, we include predictions with a 
10\arcsec$\times$10\arcsec\ FWHM convolution sampled 
near the peak surface brightness positions. The exact choice of 
kernel size is not important, but is chosen to 
combine several neighboring fibers. 
Finally, we include 
a prediction for the average surface brightness of all 
fibers expected to sample $\mu>10^{-19}$ erg/s/cm$^2$/\sq\arcsec. These 
various predictions will be compared to co-added data 
in \S \ref{sec_data}. 
We give in Figure \ref{fig_MC} the HI fits along major and minor axis cuts. The 
fits to UGC~7321 use all the HI data beyond an inner radius cut, 
which was chosen to avoid a substructure bump near $R\approx140\arcsec$. 
The fits to UGC~1281 are more constrained with both an inner and 
outer radius cut. The outer cut is to exclude a known $\sim8\arcdeg$ warp 
\citep{Gar02}. The fitting function assumes a single position angle at all 
radii and does not describe warps well. We have investigated the 
disk's outer behavior by also deriving fits from the R$>$220\arcsec\ 
data alone. A radial scale length compatible with, but noisier than, the 
Table \ref{tab_MC} value was found with a significant change in 
position angle, a reflection of the warp. 

\subsection{HI bounded limit}
\label{sec_alt_lim}
\par While we cannot definitively prove that these galaxies maintain their 
extrapolated hydrogen profiles over the galactocentric distances we will 
discuss in \S \ref{sec_mod_fit}, such an assumption is reasonable 
with the current 21 cm data. There is no evidence for flaring in these galaxies, and the 
superthin shape implies an undisturbed history. In UGC~7321, \citet{Uso03} have 
searched for low-mass compansions and found none to the limit of 
M$_{HI}=2.2\times10^{6}$ M$_{\odot}$ within 12\arcmin (35 kpc). The nearest optical 
companions are two dwarf galaxies at minimum distances of 340 kpc, implying 
minimum times to last encounter of $1.6\times10^{9}$ years. So, it is 
unlikely that gas has been stripped from the regions over which 
we have extrapolated a density profile. However, we have calculated an 
alternative limit using data bounded by the HI data in a manner 
similar to the analysis in \citet{Sto91,Vog95,Don95,Wey01} as an 
alternative, which is equivalent to assuming that the gas is completely 
truncated where the 21 cm signal falls below the noise. 
In those works, a single, simple equation based on global photoionization 
equilibrium is used and here repeated in Equation \ref{eq_projarea}. 
\begin{equation}
\label{eq_projarea}
\Phi=\Gamma\frac{3+\beta}{4a_{\nu}\beta}=\frac{I_{H\alpha}}{f_{a}f_{H\alpha}} \frac{A_{proj}}{A_{tot}}
\end{equation}

The variable $\Phi$ is the one-sided incident ionizing UVB flux in units of cm$^{-2}$ s$^{-1}$, 
$I_{H\alpha}$ is the H$\alpha$ surface brightness 
in units of $\mu$R, $f_{a}$ is the 
fraction of incident photons that become absorbed when passing through the face-on cloud, 
$f_{H\alpha}$ is the fraction of excited recombinations that produce an H$\alpha$ photon, 
$A_{proj}$ is the projected area covered by spectroscopy and 21cm data, and 
$A_{tot}$ is the total surface area for the regions in projection that can 
absorb Lyman limit photons. The area aspect ratio is usually determined 
from 21cm data. This calculation takes no account of the spatial stratification 
between 21cm and H$\alpha$ that can realistically occur for very thin 
gas distributions, as we will see later in \S \ref{sec_mod_fit} where the 
predicted H$\alpha$ surface brightness is derived, and 
requires H$\alpha$ searches and interpretations to be restricted to 
area covered by deep 21cm data. However, for mild aspect ratios ($\sim <10$) 
or large 21cm beams, this method delivers similar predictions as those in \S \ref{sec_mod_HI}. 
\par We now discuss the evaluation of the few terms in this model. The assumption in the 
HI bounded limit is that the hydrogen resides within some well-defined 
area represented by the noise floor of the 21cm data. It is not 
obvious how the area should be defined in a continuous gas distribution, but 
we adopt the photoionization front we have previously defined in Equations \ref{eq_rc} 
and \ref{eq_zc} as a realistic edge. In \S \ref{sec_mod_fit} we determine gas geometries for our target galaxies. 
In particular for the area in UGC~7321 covered by fibers, with $N_{HI}>10^{19}$ cm$^{-2}$, 
and the parameters in Table \ref{tab_MC}, we find 
$\langle\frac{A_{tot}}{A_{proj}}\rangle=24.8^{+3.4}_{-1.5}$. This value is in good agreement 
with the 21cm axis ratio of 29 determined at the $10^{20}$ cm$^{-2}$ contour 
in \citet[][Table 3]{Uso03}. By adopting this 
distribution in face-on column density and a UVB spectral index of $\beta=1.8$, 
we can evaluate $f_{a}$. We find $\langle\frac{A_{tot}f_{a}}{A_{proj}}\rangle=22.8^{+4.4}_{-1.8}$ in 
UGC~7321. With the same calculations applied to UGC~1281, we find 
$\langle\frac{A_{tot}}{A_{proj}}\rangle=19.0^{+5.6}_{-1.8}$ and 
$\langle\frac{A_{tot}f_{a}}{A_{proj}}\rangle=13.6^{+6.2}_{-2.0}$. 
Identically to \citet{Wey01}, we adopt $f_{H\alpha}=0.45$ as appropriate for case B and a 
$10^{4}$K electron temperature. We also carry out this analysis in Tables 
\ref{tab_MC} and \ref{tab_lims}  
for continuity with previous work, but we emphasize that our preferred limit comes 
from the comparisons to the model in \S \ref{sec_mod_HI} as it incorporates the 
spatial segregation between the brightest H$\alpha$ regions and the HI 
data that is natural in very thin, edge-on geometries. 

\section{H$\alpha$ data and analysis}
\label{sec_data}
\par We have obtained new integral field spectroscopy 
positioned along the major axes of UGC~7321 and UGC~1281 targeting 
H$\alpha$ with the Visible Integral-field Replicable Unit 
Spectrograph Prototype \cite[VIRUS-P,][]{Hil08a} on the McDonald 2.7m telescope. 
We observed UGC~1281 on 2009 October 22-24 with R~=~1288 from 4700-6990\AA\ for 
21 photometric hours and UGC~7321 on 2010 April 9 and 11 with a resolution of 
R~=~3860 from 6040-6740\AA\ for 15 hours under non-photometric conditions. 
Between the R~=~1288 and R~=~3860 observations, made 
possible by a new grating, we not only gain in sensitivity scaled by 
the square root of the resolution but resolve the bright skylines, 
OH $\lambda$6568.779 and geocoronal H$\alpha$, from 
our target wavelengths. We have set the controller to bin pixels by two 
in the wavelength direction which samples the spectra just at the Nyquist 
criterion and minimizes read noise.  
The VIRUS-P field covers a 1\farcm6$\times$1\farcm6 field with 
246 fibers of 2\farcs05 radius with a one-third fill-factor. We split our 
observations into three dithers to cover the entire field. In UGC~1281 
we split our time further between two overlapping fields to 
cover the outer plane better in the presence of a possible 
$<$8\arcdeg\ warp \citep{Gar02} yielding a total of six dithers. 
Spectrophotometric flux standard stars from \citet{Mas88} 
were measured once or twice nightly. We 
tracked the transparency through the offset guiding 
camera. Galactic extinction corrections \citep{Sch98,Odo94} were 
made with A$_{\mbox{V}}$=0.09 and A$_{\mbox{V}}$=0.15 for 
UGC~7321 and UGC~1281 respectively. 
A spectral airmass/extinction curve specifically modelled for the McDonald 
Observatory site was applied. We estimate its 
systematic uncertainty by comparing it to the Kitt Peak curve supplied 
with the IRAF package {\tt onedspec}. We find 
a 20\% rms difference between the wavelengths of 6000-7000\AA. The two curves 
deviate systematically at $\lambda > 5900$\AA. We believe the site specific 
McDonald curve to be more accurate to our data. However, we 
propagate the difference as a potential, systematic uncertainty. The flux 
calibration uncertainty due to the airmass/extinction curve at the 
data's median airmass of 1.09 is $\pm$0.023 magnitudes. 
\subsection{Flux calibration}
\label{sec_flux}
\par The 8\arcmin\ offset guiding camera is an Apogee Alta with a 20.25$\sq\arcmin$ 
field-of-view under a B+V ($\lambda_{\mbox{mean}}=5000$\AA) filter. 
Guider images were read out and saved every few seconds. Stacks 
of guider images that overlapped in time with each 
individual VIRUS-P exposure (20 minutes each on UGC~1281, 
30 minutes each on UGC~7321, and 1 minute 
each on the flux standards) were combined. We make a relative photometry 
correction to each science frame based on the stack of guider 
images taken simultaneously with the VIRUS-P data. Typically, 
ten stars per guider frame were available for photometry. 
\par We have switched from the standard stars to the science 
targets with gaps of less than 5 minutes and assumed 
the conditions to be constant over that time and between 
the standard star and galaxy positions to make the 
absolute flux calibration. The 
observations of standard stars were taken during the 
most photometrically stable periods during each 
night to mitigate this potential source of error. Even so, the final flux 
calibration factor we apply may have systematic 
errors. We assess 
this error by considering the 5 observations of 
2 standards, PG1708+602 and Feige 34, taken along 
with the UGC~7321 data and the 
3 observations of 1 standard, Feige 110, taken along with the 
UGC~1281 data. The distribution 
in flux calibrations is wavelength-independent over our 
observed range with a 6.8\% rms and 
2.2\% rms respectively. These estimates also capture possible
variation in transparency with on-sky position. They are reported 
in Table \ref{tab_lims} along with the possible error in the extinction curve 
between the effective wavelength of the guider and the wavelength of H$\alpha$. 
For the non-photometric 
data on UGC~7321, we measured 
a median zeropoint change, $\Delta\mbox{zp}$, of 0.276 magnitudes 
and a 68\%\ range of 0.171-0.382 magnitudes over the two nights. The more 
nearly photometric data on UGC~1281 had median $\Delta\mbox{zp}=$0.057 
magnitudes and a 68\%\ range of 0.043-0.077 magnitudes over the three nights.
\subsection{Sky background subtraction}
\par The choice of sky subtraction is particularly important for 
this work which reaches for flux limits far below the average sky 
brightness. If the science field were covered with source emission, 
sky nods would be necessary. Then, the time variability of the 
OH and geocoronal H$\alpha$ sky lines would form important 
systematic error sources. Fortunately, the large VIRUS-P field of view and 
selection of extremely thin, edge-on target galaxies affords a subset of 
fibers that contain a negligible amount of source flux to serve as simultaneously 
measured sky fibers. We selected 
fibers sufficiently far from the major axis such that the models 
predicted $\mu < 2 \times 10^{-21}$ erg/s/cm$^2$/\sq\arcsec 
(with the baseline $\Gamma=4\times 10^{-14}$ s$^{-1}$), or 100$\times$ 
below the expected peak surface brightness, 
to be used for sky subtraction. This cut left 24\% and 44\% of the fibers for 
sky estimation in UGC~7321 and UGC~1281 respectively. We 
experimented with moving this sky fiber cut up and down by a factor of five and found 
no difference in the final upper limits to the UVB strength. Depending on the number of fibers co-added, the statistical 
H$\alpha$ flux errors presented here reach to 300$\times$ dimmer than the sky level. Without simultaneously 
measured sky background, the systematics of sky nods would quickly dominate the limits. 
\subsection{Data reduction}
\par The data reduction, optimal background subtraction, and search for emission lines were completed with 
algorithms developed for a 
Lyman-$\alpha$ emitter survey \citep{Adam10a}. We summarize here the 
important steps. First, overscans and a master bias frame are subtracted from each frame. 
The wavelength solution for each fiber is fit as a fourth order polynomial to $\sim30$ emission lines from 
HgCd lamps passing through the entire telescope light path. The residuals to the solution are 
of order one hundredth of a resolution element. Flat fields 
precise to $<1\%$ are made from twilight flats with the solar spectrum removed by a b-spline fit \citep{Die93} 
and division. This fit method is the same as we apply to fitting and subtracting the sky background and 
has important advantages over data interpolation. By avoiding data resampling, we keep the errors largely 
uncorrelated. Small distortions of the instrument camera over a regular pixel grid lead to the 
spectrum from each fiber being sampled at slightly different wavelengths. By considering a collection of 
fibers together in a fit, the spectrum is oversampled, and we can recover nearly blended features. 
This method delivers an optimal spectral model robust against cosmic rays and without the residuals that linear 
interpolation can create. A thorough description of b-spline fits as applied 
to astronomy datasets can be found in \citet{Kel03}. The next step in the data reduction is to fit and 
subtract a b-spline sky background modelled from selected sky fibers. Next, cosmic rays are masked 
by finding all pixels that deviate from the other pixels in the same fiber by 
some large threshold value. Some dim cosmic rays are missed by this step, but are rejected when  
combining multiple frames. We have chosen a threshold that misses the weakest $\sim$20\% of cosmic rays 
for direct masking in this work. The exact threshold does not affect the results. The frame is then flux calibrated with the 
non-photometric zeropoint correction and airmass correction applied. Finally, a one dimensional final spectrum for 
each fiber position is created by combining all the frames taken at the same dither position and running 
across the 5 pixel cross-dispersion aperture. For the final estimate to be immune to 
remaining cosmic rays we have used the biweight estimator \citep{Bee90} at this step.
Our pipeline makes no cross-talk correction 
since we restrict our cross-dispersion apertures to 5 pixels where the fiber 
separations are typically 8 pixels and the cross-dispersion FWHMs are typically 4 pixels. This leads to, at most, 
$10\%$ contamination from neighboring fibers and becomes especially trivial when considering large collections 
of fibers as an aperture. The scattered light properties of the instrument have been characterized in 
\citet{Ada08} and, particularly at H$\alpha$ wavelengths, no scattered light or ghost patterns are found. The spectral resolution 
varies by $<5\%$ for all fibers at a common wavelength due to careful design and alignment of the 
spectrograph camera. We have made no corrections by convolution to a common resolution. The 
effect of the spectral resolution variation and the background subtraction scheme is 
to leave residuals under bright skylines. We characterize the spectral resolution 
systematic in \S \ref{sec_lim}. 
Given the large number of independent spectral elements in
VIRUS-P data (126,000 in each dither), we must choose a high 
significance cut. At 5$\sigma$ significance, the chance of noise 
leading to a detection at a particular wavelength in a 
particular dither is only 1 in 14,000. We choose to quote this 
limit as sufficiently conservative. 
\subsection{Emission line detection}
\par We describe here an automated emission line search algorithm to work with a sky 
background and continuum subtracted spectrum or stacks of spectra. By 
applying this search, we robustly find all significant emission lines at all 
redshifts. In practice, we find no significant H$\alpha$ emission with plausible 
velocity offsets in any fiber for either galaxy. 
Plausible velocity offsets are determined by the HI rotation curves.
In UGC~7321, for example, the rotation curve is flat over our data range
with variations of only $\pm$10 km s$^{-1}$. The gas dispersion
is measured in the HI data to be near 7 km s$^{-1}$ subject to 
the limitation of the 5 km s$^{-1}$ resolution 
\citep{Uso03}. Over a very conservative $\pm$100 km s$^{-1}$ (2.2\AA) range
around our target wavelengths, the flux limit is flat.
First, spectral pixels at any wavelength that exceed the noise by  
1$\sigma$ are treated as seeds. Around each seed, we fit Gaussian profiles 
of variable intensity, width, and central wavelength. The S/N of an emission line is then 
calculated by summing all pixels and errors in quadrature within 
$\pm$2$\sigma_{res}$ for the wavelength of interest where
$\sigma_{res}$ is the instrumental dispersion. 
In the
UGC~7321 data, $\sigma_{res}=$33 km s$^{-1}$, and in the UGC~1281 data,
$\sigma_{res}=$100 km s$^{-1}$.
When quoting limits on undetected emission lines, we sum in quadrature the errors within 
the same spectral window. These steps in error combination consider both the 
statistical errors in the reduced data and the systematic 
error based on ill-matched spectral resolution between fibers discussed in \S \ref{sec_err}. A 
spectral correction factor is divided into the detections and limits to consider 
the fraction of a Gaussian function's flux that falls outside of the considered window 
as $f_{spec}=\mbox{erf}(\sqrt{2}\sigma_{res}/\sqrt{\sigma_{res}^2+\sigma_{det}^2})$ where 
$\sigma_{det}$ is the detected emission line width. 
This same factor determines the degradation in flux limit for broad line
detections. 
For unresolved limits, $\sigma_{det}$ is 
considered to be zero and the spectral correction ($f_{spec}^{-1}$) evaluates as $\times 1.05$. 
In practice, we make no significant detections within $\pm$500 km/s of the 
HI based expected velocity in either galaxy. The average HI 
heliocentric velocities of UGC~7321 and UGC~1281 are 407 km/s \citep{Uso03} and 157 km/s \citep{Gar02}
with the asymptotic HI velocities nearest our pointings at 
$\sim$510 km/s and $\sim$210 km/s respectively. We observed 
under topocentric radial velocities of -12 km/s and 3 km/s 
toward UGC~7321 and UGC~1281 respectively. Therefore, we expect unresolved H$\alpha$ 
emission at 6573.7$\pm$0.4\AA\ and 6567.5$\pm$0.8\AA\, using the 
asymptotic values just quoted, in the observed frames of 
UGC~7321 and UGC~1281 respectively. The gas velocity dispersions in the 
21 cm data are of the order 7 km s$^{-1}$. The 
21 cm rotation curves change by $\pm10$ km s$^{-1}$ over our fields. 
These two values form the expected wavelength range, and the flux 
limits around these lines are flat to $\pm100$ km s$^{-1}$. 
\par Background galaxies produce the dominant flux in a number of fibers. 
This is evident where we can measure redshifts through emission lines identifiable as 
either Lyman-$\alpha$, [OII]$\lambda$3727,
H$\beta$, [OIII]$\lambda$4959, or [OIII]$\lambda$5007. For most of the background 
systems with emission lines the redshift is determined by the pattern of multiple emission lines. 
If the background galaxies have smooth continuum through our wavelength of interest their 
removal is accomplished in the continuum removal step. However, the possibility 
of spectral structure in the continuum across the corresponding H$\alpha$ wavelength range leads us to 
mask those regions. 
Operationally, we mask a fiber if it displays a 5$\sigma$ significant 
value in its continuum as estimated across all available wavelengths under 
inverse variance weighting. It is also possible that 
weak continuum is coming from objects in the halo of the target galaxies, in 
which case the desirability of a mask is less certain. We have performed all the 
emission line searches and limits with and without this masking process and found 
no detections in either case. The values we present as limits were made with the 
masks applied. 
\subsection{Data co-addition and limits}
\label{sec_lim}
\par
\begin{figure}
\centering
\includegraphics [scale=0.8,angle=0]{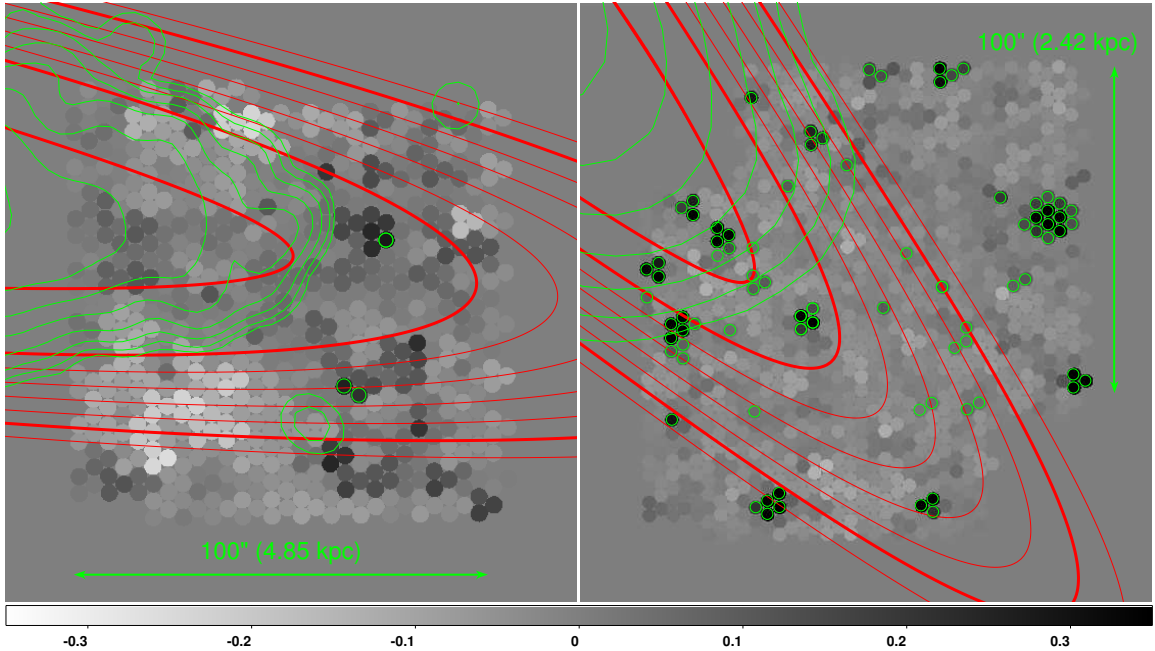}
\caption{\textit{Left} Reconstructed VIRUS-P continuum 
image of the 
UGC~7321 outskirts centered at $\alpha_{J2000}$=12:17:16.4 
and $\delta_{J2000}$=+22:31:33 or
$\approx$250$\arcsec$ off the minor axis. The continuum estimation is 
made through the entire available spectral range from 6040-6740\AA\ 
with the colorbar in units of $10^{-17}$ 
erg s$^{-1}$ cm$^{-2}$ \AA$^{-1}$. The dark, circled 
objects are masked as background galaxies, many known to be 
background by their emission lines at redshifts higher than 
the target galaxy's redshift. 
One can see some broad structure in the continuum map due to small 
residuals in the fiber-to-fiber throughput as 
described in \S \ref{sec_err}, especially in the UGC~7321 data. 
The green 
contours trace the HI column densities in steps of 
(10,19,36,67,126,238,448,845) 
$\times 10^{18}$ cm$^{-2}$. The red, more extended contours trace the 
predicted H$\alpha$ surface brightness assuming 
$\Gamma=4\times10^{-14}$ s$^{-1}$ and $\beta=1.8$ in contour levels of  
(0.1,0.24,0.57,1.4,3.3,7.9,19) 
$\times 10^{-20}$ erg/s/cm$^2$/\sq\arcsec. The 
two innermost red contours enclose the surface 
brightness maxima. Positions 
closer to the center again become dimmer in H$\alpha$ since 
portions of the gas, in projection, stay neutral at smaller radii. 
The fibers used in sky 
subtraction are all those outside the second outermost
red contour. We draw the second, seventh, and eighth contours thickly to 
highlight these regions. As a scale reference, the fiber diameter is 4\farcs1. 
 \textit{Right} The same display for UGC~1281 with central position 
$\alpha_{J2000}$=1:49:15.8 and $\delta_{J2000}$=+32:31:46 or 
$\approx300\arcsec$ off the minor axis. The continuum estimation is
made through the entire available spectral range from 4700-6990\AA. Here, many 
more background galaxies are found. In UGC~1281, we took data at two 
overlapping fields. The central positions covered by both square pointings have the best depth.
}
\label{fig_image}
\end{figure}

 We show the derived limits in Table \ref{tab_lims}. 
We find no significant emission lines within the vicinity of the galaxies' 
velocities (defined as within $\pm500$ km s$^{-1}$) in 
any individual fiber. We next 
mask out continuum sources and 
apply a circular spatial filter as a 2D Gaussian function kernel with 
FWHM=10\arcsec. Again, 
we find no significant emission. Finally, 
we stack all fibers for which the model of \S \ref{sec_mod_HI} predicts $\mu > 10^{-19}$ 
erg/s/cm$^2$/\sq\arcsec. 
The choice of the cut in $\mu$ is not rigorously
determined, but judged as sensible from the
shape of the surface brightness distribution in Figure \ref{fig_image} 
which indicates that many dozens of fibers contain predicted
surface brightnesses at roughly one-half the peak value. The
co-addition of these fibers to one peak fiber will obviously yield an
improved S/N. Different choices in the cut will lead to slightly different
formal limits, but the fractional effect is small once large
stack sizes of several hundred fibers are reached. 
The models 
used to select those fibers are those presented in Table 
\ref{tab_MC} with an assumed $\Gamma=4\times10^{-14}$ s$^{-1}$ 
and $\beta=1.8$. We use the nearly (Figure \ref{fig_lintest}) linear scaling between 
$\Gamma$ and $\mu$ to determine the true value of $\Gamma$. 
The models predict such averages to yield 
1.7$\times 10^{-19}$ erg/s/cm$^2$/\sq\arcsec\ for UGC~7321 and 
1.8$\times 10^{-19}$ erg/s/cm$^2$/\sq\arcsec\ for UGC~1281. We again 
find no significant emission in the stacked spectra. These 
emission line searches were performed solely with errors based on Poisson 
noise statistics and yielded no detections. In \S \ref{sec_err} we 
discuss additional systematic errors that degrade the 
final limits derived from purely Poisson errors in the data. By the models, the peak H$\alpha$ 
surface brightness would 
have fallen in our fields for UVB strengths from 
2$\times 10^{-14}$ s$^{-1}<\Gamma<$ 2$\times 10^{-12}$ 
s$^{-1}$ and warps of $<$12.4\arcdeg\ in 
UGC~7321 and 4$\times 10^{-15}$ s$^{-1}<\Gamma<$ 3$\times 10^{-13}$ 
s$^{-1}$ and warps of $<$15.8\arcdeg\ 
in UGC~1281. However, a radial displacement of the field would 
still give significant flux as seen in the contour plots, so we do 
not expect misalignments to affect the final limits. Figure 
\ref{fig_image} shows the positions of the observations 
relative to several key features. The 21 cm data contours 
are overlayed, the locations of masked background 
galaxies are shown, and the expected spatial 
profiles of H$\alpha$ emission is shown. 
We show in Figure \ref{fig_UGC7321_spec} the sky spectra and the three types of 
spectral stacks to background subtracted data in UGC~7321. In 
Figure \ref{fig_UGC1281_spec} we show the corresponding ones for UGC~1281. In 
neither case do we make a significant detection in H$\alpha$. 
\par Our selection of fibers for co-addition based on an assumed value of $\Gamma$ 
leads us in turn to a lower limit on $\Gamma$. This may in principle introduce 
an error into our determination of $\Gamma$. However, both a rough estimation and then a detailed 
analysis show that the non-linearity in this operation is negligible. 
First, one measure of the spatial scale of the H$\alpha$ surface brightness profile
is the threshold radius, $r_c$, of Equation \ref{eq_rc}. Since $r_c$
scales only as the natural logarithm of $\Gamma$, there is little
change over the range of possible UVB strengths that we consider.
The shape of the H$\alpha$ surface brightness profile is also
broad and smooth, from Figure \ref{fig_image}, relative to the possible
range of $r_c$. By selecting wide swaths of fibers for co-addition, the
problem is particularly well behaved. Second, we verify these
arguments with a numerical example. We simulated the
surface brightness profiles for $\Gamma=8\times10^{-15}$ s$^{-1}$,
or five times lower than the nominal modeled value. We sampled the
same set of fibers for co-addition as with the previous analysis.
The model, average surface brightness was
$\mu=3.1\times 10^{-20}$ erg/s/cm$^2$/\sq\arcsec\ and
$\mu=3.9\times 10^{-20}$ erg/s/cm$^2$/\sq\arcsec\
for UGC~7321 and UGC~1281 respectively, or only
6\% lower and 10\% higher than the linear prediction. We conclude
that the selection of co-added fibers based on the
nominal UVB strength has negligible impact our final limit.
\subsection{Error assessment}
\label{sec_err}
\par There are several potential sources of systematic error to the 
presented spectra. We have already discussed the uncertainties 
in the model-based conversion of H$\alpha$ surface brightness to 
UVB strength in \S \ref{sec_mod_fit}. The uncertainty in the 
absolute spectral flux calibration due to the applied 
atmospheric extinction curve is discussed in \S \ref{sec_data}. 
The uncertainty in the absolute spectral flux calibration due to the 
standard star observations is discussed in \S \ref{sec_flux}. We now 
analyze a final systematic regarding the relative error 
determinations in the H$\alpha$ spectra. We observe that the propagation 
of the errors from the data's original read noise and 
shot noise does not fully account for the variation in sky subtracted 
spectra. This is especially true under 
bright skylines. We discuss three possible causes with a focus on 
the variation of spectral resolution across different fibers. In any 
of the cases, the form of the systematic error will be to add a small 
percentage of the continuum subtracted sky background spectrum applied linearly 
with the random error. 
\par
\begin{figure}
\centering
\includegraphics [scale=0.45,angle=-90]{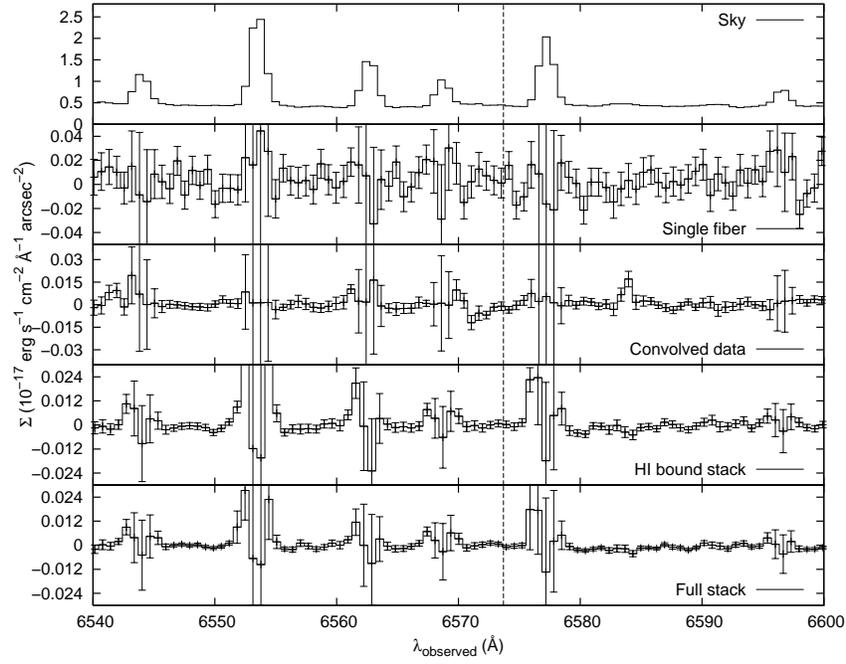}
\caption{Selected spectra around H$\alpha$ in UGC~7321 presented 
in units of surface brightness. The 
expected wavelength for emission is shown with a dotted line. 
The frames from top to bottom show the background 
sky, the background subtracted spectrum for a 
typical fiber that does not display continuum, the 
spectrum at the same position after being smoothed 
by a 10\arcsec\ FWHM circular Gaussian kernel, the data 
bounded by HI signal, and 
finally the stack of the 358 fibers 
predicted to be the brightest by the model. The 
errorbars consist of the Poisson, observational 
error and the systematic spectral resolution error 
of columns two and three in Table \ref{tab_lims} only. 
The spectral resolution systematic, discussed in \S \ref{sec_err}, 
is most important under the bright skylines and does not 
dominate at the target wavelength. 
}
\label{fig_UGC7321_spec}
\end{figure}

\begin{figure}
\centering
\includegraphics [scale=0.45,angle=-90]{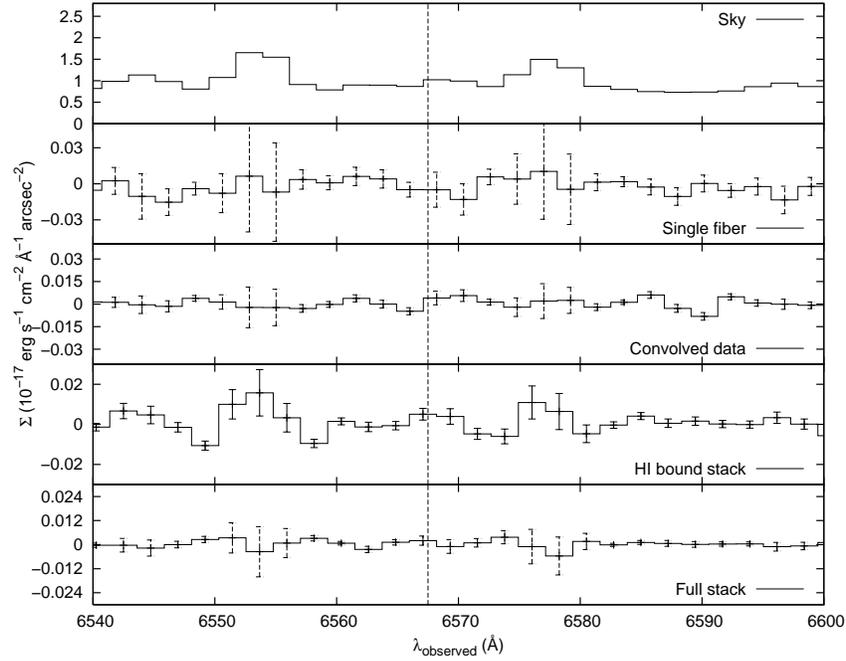}
\caption{Selected surface brightness spectra around H$\alpha$ in UGC~1281. 
The format is the same as in Figure \ref{fig_UGC7321_spec}. In this 
case, 313 of the brightest expected fibers form the final stack.  
}
\label{fig_UGC1281_spec}
\end{figure}

 First, the instrumental spectral resolution varies by at most 5\% in different 
fibers due to small but detectable optical distortions in the camera. We further measure 
from arc lamp exposures that the variation is 2.5\% between 
the sky and science fibers in the 
UGC~7321 data and 1.5\% in the UGC~1281 data. These factors 
are presented in column 3 of Table \ref{tab_lims} and 
scaled by the background subtracted sky spectrum and applied as systematic 
errors in the spectra presented in Figures \ref{fig_UGC7321_spec} and \ref{fig_UGC1281_spec}. This form 
of the systematic, as the fractional error in the dispersion times the background 
subtracted sky spectrum, can be derived simply by taking the first order 
expansion of a Gaussian function near the line center. Second, 
the fiber-to-fiber throughput can vary slightly between flat field calibrations. 
The relative fiber-to-fiber throughput is calibrated with sky flats 
taken at dawn and dusk. This relative throughput has been measured to 
be stable to $<$5\% over most nights. However, we find a maximum 
15\% fiber-to-fiber throughput variation in the UGC~7321 data 
due to poor fiber cable coiling practices. This error is very 
evident in the broadband estimate per fiber as shown in 
Figure \ref{fig_image}. The error is less important for a 
continuum subtracted spectral element where most of the 
fiber-to-fiber throughput error subtracts out. The form of the 
throughput variation is that a few fibers experience a change with time, but 
the majority stay stable. We measure the 
rms throughput variation between all fibers to be far below 
1\%. Third, sky lines may vary across the $\sim1\arcmin$ separating 
the sky and science fibers. The UGC~7321 fiducial ``signal'' is well resolved from all 
known sky lines and only near OH lines, but the UGC~1281 ``signal'' is 
unresolved from an OH line and near the geocoronal H$\alpha$. 
Variations on such small spatial scales have not been observed, and the data are 
averaged over very long integration times and large ranges 
in zenith distance, so we do not expect sky variation over our field-of-view to be a dominant 
error term. It is possible that the 
geocoronal H$\alpha$ emission may vary within $\sim1\arcmin$, 
but small-scale variation is less likely for OH. We choose to 
parameterize the total effect of these systematics in a conversative 
manner by deriving from the data themselves the systematic error based on the 
measured levels of spectral resolution variation seen between fibers. 
\par This systematic error strongly affects the UGC~1281 data since 
the lower resolution allows blending of night sky lines at the 
expected wavelength of H$\alpha$, but it is a less 
important component to the UGC~7321 error budget. As data from more fibers are 
coadded, this systematic error takes on greater importance in relation to the 
random error. 
We assess the $\chi^2$ distributions across 6300-6600\AA\ in each 
co-addition case in Table \ref{tab_lims}. The 
$\chi^2$ are simply calculated against a flat, zero flux line and 
can be visually judged in Figures \ref{fig_UGC7321_spec} and
\ref{fig_UGC1281_spec}. 
The distributions look 
very symmetric around zero, and the reduced $\chi^2$ values 
are consistent with noise. The proper $\chi^2$ values 
validate our systematic noise estimates empirically. In fact, the 
additional noise estimates may be slightly conservative. One can visually 
note from Figures \ref{fig_UGC7321_spec} and \ref{fig_UGC1281_spec} 
that the $\chi^2$ values are even lower than the degrees of freedom ($\nu$) 
in the most important regions near the target wavelengths. 
\subsection{Internal galactic extinctions}
\par Internal extinctions in disk galaxies at these 
scale lengths are very uncertain despite being a subject 
of detailed research \cite[e.g.][]{Byu94}. \citet{Mat99} see in 
UGC~7321 an abrupt 
truncation of resolved dust clumps beyond r$\approx$80\arcsec\ 
and fit a model of radially declining dust where, for our 
position around 250\arcsec, 
there is no internal extinction. We have taken short VIRUS-P 
exposures covering H$\alpha$ and H$\beta$ on the 
galaxy centers to derive conversative internal extinction 
upper limit estimates after correction for Galactic extinction. 
We did not take deep enough exposures to measure accurate 
stellar populations and photospheric Balmer absorptions ourselves, so we 
have relied on literature values appropriate to late type galaxies. 
From the Balmer decrements we measure 
A$_{H\alpha}$=-0.03$\pm$0.09 magnitudes for UGC~7321 and 
A$_{H\alpha}$=-0.02$\pm$0.11 magnitudes for UGC~1281 
under the assumption that the absorption equivalent widths 
satisfy EW(H$\alpha$)$_{abs}$=EW(H$\beta$)$_{abs}$=2\AA\ 
\citep{McC85,Cal94}. 
As the extinction estimates are consistent with zero, we apply no dust 
correction to our results.
\section{Discussion}
\label{sec_dis}
\par The flux decrement method is currently the most widely used method to 
estimate the UVB strength at high redshift. Under the fluctuating Gunn-Peterson 
approximation \citep{Cro98}, the Lyman-$\alpha$ forest optical depth 
distribution should have a normalization that depends only on well constrainted 
cosmological parameters and the UVB strength. The IGM temperature and density 
distributions may have some systematic uncertainties that propagate into 
knowledge of the UVB, but they are not likely the leading uncertainties. 
The more likely dominant uncertainties in flux decrement modeling are the source 
emissivities. At $z\lessapprox1$, the Lyman limit mean free path becomes larger than the horizon, so 
the UVB strength at z=0 is influenced by source evolution across this redshift range. 
AGN and stellar population luminosity functions, both observed and 
modeled, generally agree to better than an order of magnitude over these 
redshifts. The least constrained input to flux decrement modeling is the 
escape fraction for ionizing photons in galaxies, particularly at low redshift and 
low luminosity. We believe our measurement is best interpreted as an indicator of a low 
escape fraction.
\par
\begin{figure}
\centering
\includegraphics [scale=0.55,angle=-90]{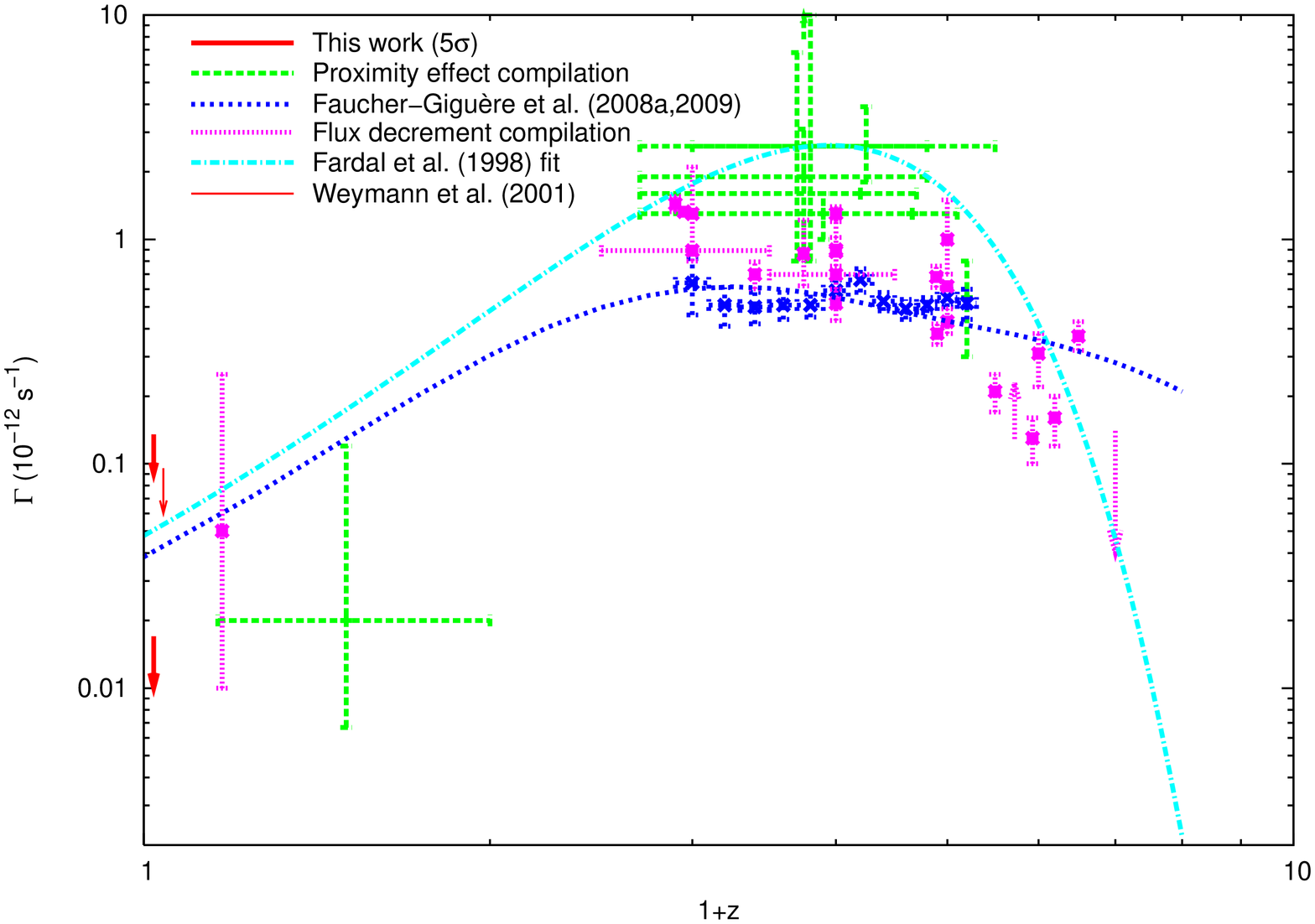}
\caption{A compilation of photoionization rates across redshift. 
Most of the literature 
compilations come from Table 2 in \citet{Fau08c} and Table 1 in \citet{Fau08a}. 
The flux decrement measurement at $z\sim0.17$ is from 
\citet{Dav01}. The low redshift, H$\alpha$ limit from 
\citet{Wey01} (2$\sigma$) has been the deepest z$=$0 limit 
before this work. The UVB fitting function comes from 
\citet{Far98} and the newer simulation from \citet{Fau09}. Our 
work's new limit 
is well below the flux decrement normalized simulation and 
challenges one or more of the model assumptions. Some points have 
been slightly shifted in redshift for visual clarity.
}
\label{fig_zGam}
\end{figure}

 Our most constraining (5$\sigma$) spectral limits are 
$\Gamma<1.7\times 10^{-14}$ s$^{-1}$ in UGC~7321 and $\Gamma<13.5\times 10^{-14}$ s$^{-1}$ 
in UGC~1281 again assuming $\beta=1.8$. Several benchmarks, both empirical and 
theoretical, exist with which to compare these limits. Figure \ref{fig_zGam} shows
the UVB strength against redshift determined by many groups.  
The lowest redshift proximity effect limit comes from 
\citet{Kul93} with analysis of 13 quasars from \citet{Bah93} between 
$0.16 \le z \le 1.00$ at $\Gamma(\bar{z}=0.5)=2.0^{+10}_{-1.3}\times 10^{-14}$ s$^{-1}$. 
However, the proximity effect method has been shown to have a high bias that 
depends on halo mass \citep{Fau08c} and should be 
interpreted with care. 
The theoretical model of \citet{Fau09} gives a drop in the UVB strength by a 
factor of 3.4 between z=0.5 and z=0.0 leaving this measurement 
consistent with our current limit. This agreement is interesting and somewhat 
unexpected given the bias of proximity effect measurements. The only 
existing low-z flux decrement limit is 
$\Gamma(\bar{z}=0.17)=5.0^{+20.}_{-4.0}\times 10^{-14}$ s$^{-1}$ 
\citep{Dav01}. The theoretical model itself, normalized 
by the flux decrement method, 
predicts $\Gamma(z=0)=3.8\times 10^{-14}$ s$^{-1}$ which is much higher than 
our new limit. There exists a second set of unpublished theoretical 
predictions from F. Haardt and P. Madau discussed in \citet{Fau09} giving 
$\Gamma(z=0)=1\times 10^{-13}$ s$^{-1}$. The latter model 
used a constant 10\% escape fraction of ionizing photons and an 
unspecified star formation history while the former used a 
completely theoretical and simulation-based star formation 
history \citep{Her03} and a scaling 
of the stellar UV emissivity based on high redshift flux decrement 
measurements that contains the escape fraction. A 
comparison to Lyman-break galaxy (LBG) luminosity functions led that 
group to require only $f_{esc,abs}\approx0.5\%$ \citep{Fau08a}. The 
direct measurement of galactic escape fractions is difficult due to 
the low values involved. While UV bright samples can range up to 
$\approx 3\%$ in absolute Lyman limit escape fraction \citep{Sha06}, 
a presumably lower-mass sample yielded 
$(2\pm2)\%$ \citep{Che07}. Theoretical work shows a strong 
decrease in $f_{esc}$ with star formation rate and halo mass \citep{Gne08} below 
$M_{tot}\approx10^{11}M_{\odot}$, 
and lower redshift observations of populations similar to LBGs 
show a potential redshift evolution \citep{Sia10} with $f_{esc,abs}<0.8\%$. 
There is no reason yet to suppose a lower bound to the escape 
fraction. If we interpret our limit as a scaling of the escape fraction 
from the models in \citet{Fau08a} at low redshift, we find 
$f_{esc,abs}<0.2\%$. 
\par It is unlikely that systematics from the model assumptions in 
our analysis can cause the 
disagreement. Contaminating ionization from the galaxies' forming 
stars would bias our measurement high, only making the disagreement more 
severe. We further note that the degree of contamination can be 
measured by anomalous [NII]$\lambda$6548 to H$\alpha$ ratios (BFQ) 
and should not, in principle, limit this type of measurement. 
There has been a large body of work on low strength star 
formation beyond the optical radii in local galaxy disks, 
usually labelled extended UV disks (XUV), 
fostered by far UV (FUV,1350-1750\AA) and near 
UV (1750-2750\AA) Galaxy Evolution 
Explorer (GALEX) data \cite[e.g.][]{Thi07}. Narrowband 
H$\alpha$ imaging and spectroscopy have revealed that $\sim10$\% of 
gas rich disks \citep{Wer10a,Wer10b,Her10} host outlying H$\alpha$ 
emitting complexes as either compact HII regions or dwarf satellite 
companions. The common H$\alpha$ fluxes observed so far are of the order of a 
few times $10^{-16}$ erg/s/cm$^2$. Any such systems would have been found 
in our data as strong detections limited in size to a few fibers. The 
expectation of large-scale, diffuse UVB H$\alpha$ emission 
should discriminate reliably against compact XUV H$\alpha$ emission. We have also 
visually inspected the target galaxies' GALEX data which have not yet 
been analyzed in any XUV focused work. UGC~1281 
has only been covered in the rather shallow all-sky survey mode. UGC~7321 has 
been covered for 2.8ks in the NUV and 1.7 ks in the 
FUV under guest investigator 
cycle 4 proposal ID 095 (PI: J.~Lee) 
as part of the 11HUGS project \citep{Lee09}. In neither 
system is there evidence for an extended UV disk beyond the 
DSS2-red\footnote{The Digitized Sky Survey was produced at the Space Telescope 
Science Institute under U.S. Government grant NAG W-2166. The images of these 
surveys are based on photographic data obtained using the Oschin Schmidt 
Telescope on Palomar Mountain and the UK Schmidt Telescope. The plates were 
processed into the present compressed digital form with the permission of 
these institutions.} limiting contours. Finally, these contamination 
issues are speculative until a putative UVB H$\alpha$ detection is made. 
The only possible systematics that could have made a low bias to our limit 
are unaccounted for dust or gas distribution parameters, such as 
inclination, far beyond the range we have constrained. 
\par We have made our first analysis under the assumption that 
the gas distribution extends beyond the HI data limits 
with the same exponential form as at smaller radii. This assumption, 
motivated by the thin and regular HI distributions and lack of nearby companions, 
has the strongest impact on our interpretation. An alternative estimate without this 
assumption, taking only fibers that overlap with the observed HI signal, 
yields a very comparable limit of $\Gamma<2.3\times 10^{-14}$ s$^{-1}$ at 5$\sigma$ 
significance in UGC~7321. This agreement essentially comes about because our 
original model predicts only a minor H$\alpha$ contribution at the discarded 
positions under the modeled UVB strength. Nevertheless, there is no reason to 
assume the presence of an HI edge since the radio observations detect the gas 
up to the column densities where the sensitivity runs out. 
This result raises the 
question whether a redshift-dependent escape fraction is manifesting 
in galaxies. Alternatively, our new limits may be 
saying that the UVB strength, as estimated through flux decrement 
measurements, has been overestimated at all redshifts. The latter choice would upset the apparent 
agreement between current models and reionization constraints. Either case 
will require some modification to the UVB strength model and its implementation 
in structure formation simulations. We intend to pursue our measurements of these 
and other superthin galaxies to greater depth in order to arrive at a 
detection of $\Gamma(z=0)$.

\acknowledgments
We thank Karl Gebhardt, Guillermo Blanc, Benjamin Weiner, 
Jeremy Murphy, and Joss Bland-Hawthorn for fruitful 
discussion on this topic. The skills of the McDonald Observatory 
staff, and in particular David Doss, 
have been indispensable to this project. J.M.U. appreciates the support of a 
``Bourse de la Ville de Paris'' during part of this research. J.J.A. 
acknowledges the support of a National Science Foundation 
Graduate Research Fellowship and a UT David Bruton, Jr. Fellowship during this work. 
This work was partially supported by a Texas Norman Hackerman Advanced Research 
Program under grant 003658-0295-2007. 
We thank the Cynthia and George Mitchell Foundation for funding the VIRUS-P instrument. 
Finally, we thank an anonymous referee for very important improvements to this work. 
{\it Facilities:} \facility{Smith (VIRUS-P)}.

\appendix
\section{Full solution to the general H$\alpha$ surface brightness}
\label{sec_full_mu}
\par We give here the derivation of the general H$\alpha$ surface brightness at 
field positions $b_1$ and $b_2$. The special case 
for $b_1$=$r_c$ and $b_2$=0 was derived as Equation \ref{eq_SB}. That 
case is simplified since the line of sight integration can proceed 
from zero to infinity without intersecting the photoionization 
boundary and has symmetry between positive and negative values of $\rho$. 
For the general case, the simple task presented in this Appendix 
is to find the possible geometrical 
intersections of $z$ from Equation \ref{eq_z_tran} and 
$z_c(R)$ from Equation \ref{eq_zc} as a function of $\rho$ 
under inputs $i$, $b_1$, and $b_2$. This may have zero or two intersections 
labelled as $\rho_{r1}$ and $\rho_{r2}$. Once found, the general 
expression for $\mu$ then follows Equation \ref{eq_gen_SB}.  
\begin{eqnarray}
\label{eq_gen_SB}
\mu(b_1,b_2)=\frac{\xi^2 \varepsilon \alpha_{H\alpha}^{eff}n_0^2 h \nu_{H\alpha} \gamma}{\Omega} \nonumber \\
\mbox{with } \gamma=\left\{
\begin{array}{lr}
\int^{\infty}_{-\infty}\exp(-\frac{2\rho \cos i}{h_z}-\frac{2\sqrt{r_c^2+\rho^2 \sin^2 i}}{h_r}) d\rho & : \mbox{no roots in }\rho\\
\int^{\rho_{r1}}_{-\infty}\exp(-\frac{2\rho \cos i}{h_z}-\frac{2\sqrt{r_c^2+\rho^2 \sin^2 i}}{h_r}) d\rho+\\
\int^{\infty}_{\rho_{r2}}\exp(-\frac{2\rho \cos i}{h_z}-\frac{2\sqrt{r_c^2+\rho^2 \sin^2 i}}{h_r}) d\rho & : \mbox{roots in }\rho
\end{array}
\right.
\end{eqnarray}
The first necessary condition for any intersection to occur is evidently expressed in Equation \ref{eq_check1}, 
as the largest possible distance for an intersection to lie from the galaxy center is $r_c$ while the 
closest possible approach for a sight line is $b_1$. 
\begin{equation}
\label{eq_check1}
b_1<r_c
\end{equation}
The intersections in $\rho$ can be expanded into simple quadratic equations. Each of the two 
potential roots from the quadratic solution is double valued when considering intersections with 
both signs of the $z_c(R)$ surface leading to four possible roots. However, only at most two 
of the roots will be physical with the rejected two lying on extrapolations of 
$z_c(R)$ at $R(\rho)>r_c$ or $|z(\rho)|>r_c\times h_z/h_r$. The intersections 
with these surfaces lead to possible limits $\rho_{s1}$, $\rho_{s2}$, 
$\rho_{s3}$, and $\rho_{s4}$ expressed in Equations \ref{eq_lim1}-\ref{eq_lim4}. The 
most constraining limits are then the values between these four with the smallest 
absolute values leading to Equations \ref{eq_trlim1}-\ref{eq_trlim2} for the 
active limits $\rho_{l1}$ and $\rho_{l2}$. 
\begin{equation}
\label{eq_lim1}
\rho_{s1} = \frac{-b_2 \sin i - \frac{h_z}{h_r} r_c}{\cos i}
\end{equation}
\begin{equation}
\label{eq_lim2}
\rho_{s2} = \frac{b_2 \cos i - \sqrt{r_c^2-b_1^2}}{\sin i}
\end{equation}
\begin{equation}
\label{eq_lim3}
\rho_{s3} = \frac{-b_2 \sin i + \frac{h_z}{h_r} r_c}{\cos i}
\end{equation}
\begin{equation}
\label{eq_lim4}
\rho_{s4} = \frac{b_2 \cos i + \sqrt{r_c^2-b_1^2}}{\sin i}
\end{equation}
\begin{equation}
\label{eq_trlim1}
\rho_{l1} = \max(\rho_{s1},\rho_{s2})
\end{equation}
\begin{equation}
\label{eq_trlim2}
\rho_{l2} = \min(\rho_{s3},\rho_{s4})
\end{equation}
The potential intersections with $z_c(R)$ can be directly evaluated as $\rho_{p1}$, 
$\rho_{p2}$, $\rho_{p3}$, and $\rho_{p4}$ as given in Equations 
\ref{eq_root1}-\ref{eq_root4}. 
\begin{eqnarray}
\label{eq_root1}
\rho_{p1} = \frac{\left(\frac{h_r}{h_z}\right) r_c \cos i - \left(\frac{h_r}{h_z}\right)^2 
b_2 \sin i \cos i - b_2 \sin i \cos i + \sqrt{\beta_{p1}}}{\left(\frac{h_r}{h_z}\right)^2 \cos^2 i - \sin^2 i} \nonumber \\
\beta_{p1}=2\left(\frac{h_r}{h_z}\right)^2 b_2^2 \cos^2 i \sin^2 i - 2 \left(\frac{h_r}{h_z}\right) b_2 r_c \cos^2 i \sin i 
+ b_1^2 \left(\frac{h_r}{h_z}\right)^2 \cos^2 i \nonumber \\
+ b_2^2 \left(\frac{h_r}{h_z}\right)^2 \sin^2 i 
+ r_c^2 \sin^2 i + \left(\frac{h_r}{h_z}\right)^2 b_2^2 \sin^4 i - 2 \left(\frac{h_r}{h_z}\right) r_c b_2 \sin^3 i - b_1^2 \sin^2 i
\end{eqnarray}
\begin{eqnarray}
\label{eq_root2}
\rho_{p2} = \frac{\left(\frac{h_r}{h_z}\right) r_c \cos i - \left(\frac{h_r}{h_z}\right)^2 
b_2 \sin i \cos i - b_2 \sin i \cos i - \sqrt{\beta_{p1}}}{\left(\frac{h_r}{h_z}\right)^2 \cos^2 i - \sin^2 i} \nonumber \\
\beta_{p1}=2\left(\frac{h_r}{h_z}\right)^2 b_2^2 \cos^2 i \sin^2 i - 2 \left(\frac{h_r}{h_z}\right) b_2 r_c \cos^2 i \sin i 
+ b_1^2 \left(\frac{h_r}{h_z}\right)^2 \cos^2 i \nonumber \\
+ b_2^2 \left(\frac{h_r}{h_z}\right)^2 \sin^2 i 
+ r_c^2 \sin^2 i + \left(\frac{h_r}{h_z}\right)^2 b_2^2 \sin^4 i - 2 \left(\frac{h_r}{h_z}\right) r_c b_2 \sin^3 i - b_1^2 \sin^2 i
\end{eqnarray}
\begin{eqnarray}
\label{eq_root3}
\rho_{p3} = -\frac{\left(\frac{h_r}{h_z}\right) r_c \cos i - \left(\frac{h_r}{h_z}\right)^2 
b_2 \sin i \cos i - b_2 \sin i \cos i + \sqrt{\beta_{p2}}}{\left(\frac{h_r}{h_z}\right)^2 \cos^2 i - \sin^2 i} \nonumber \\
\beta_{p2}=2\left(\frac{h_r}{h_z}\right)^2 b_2^2 \cos^2 i \sin^2 i + 2 \left(\frac{h_r}{h_z}\right) b_2 r_c \cos^2 i \sin i 
+ b_1^2 \left(\frac{h_r}{h_z}\right)^2 \cos^2 i \nonumber \\
+ b_2^2 \left(\frac{h_r}{h_z}\right)^2 \sin^2 i 
+ r_c^2 \sin^2 i + \left(\frac{h_r}{h_z}\right)^2 b_2^2 \sin^4 i + 2 \left(\frac{h_r}{h_z}\right) r_c b_2 \sin^3 i - b_1^2 \sin^2 i
\end{eqnarray}
\begin{eqnarray}
\label{eq_root4}
\rho_{p4} = -\frac{\left(\frac{h_r}{h_z}\right) r_c \cos i - \left(\frac{h_r}{h_z}\right)^2 
b_2 \sin i \cos i - b_2 \sin i \cos i - \sqrt{\beta_{p2}}}{\left(\frac{h_r}{h_z}\right)^2 \cos^2 i - \sin^2 i} \nonumber \\
\beta_{p2}=2\left(\frac{h_r}{h_z}\right)^2 b_2^2 \cos^2 i \sin^2 i + 2 \left(\frac{h_r}{h_z}\right) b_2 r_c \cos^2 i \sin i 
+ b_1^2 \left(\frac{h_r}{h_z}\right)^2 \cos^2 i \nonumber \\
+ b_2^2 \left(\frac{h_r}{h_z}\right)^2 \sin^2 i 
+ r_c^2 \sin^2 i + \left(\frac{h_r}{h_z}\right)^2 b_2^2 \sin^4 i + 2 \left(\frac{h_r}{h_z}\right) r_c b_2 \sin^3 i - b_1^2 \sin^2 i
\end{eqnarray}
The comparisons to the limits $\rho_{l1}$ and $\rho_{l2}$ 
discard unphysical values in Equations \ref{eq_lastlim1}-\ref{eq_lastlim2} where the final 
limits of integration are found. 
\begin{equation}
\label{eq_lastlim1}
\rho_{r1}=\min(x \in \{\rho_{p1},\rho_{p2},\rho_{p3},\rho_{p4}\} : \rho_{l1} < x < \rho_{l2} \})
\end{equation}
\begin{equation}
\label{eq_lastlim2}
\rho_{r2}=\max(x \in \{\rho_{p1},\rho_{p2},\rho_{p3},\rho_{p4}\} : \rho_{l1} < x < \rho_{l2} \})
\end{equation}
With the integration boundaries now well defined, $\mu(b_1,b_2)$ can easily be obtained through numerical integration.

\bibliography{UVX_Ha}   

\begin{thebibliography}{72}
\expandafter\ifx\csname natexlab\endcsname\relax\def\natexlab#1{#1}\fi

\bibitem[{{Adams} {et~al.}(2008){Adams}, {Hill}, \& {MacQueen}}]{Ada08}
{Adams}, J.~J., {Hill}, G.~J., \& {MacQueen}, P.~J. 2008, in \procspie, Vol.
  7014, 232

\bibitem[{{Adams} {et~al.}(2010)}]{Adam10a}
{Adams}, J.~J., {et~al.} 2010, accepted to ApJS, ArXiv:1011.0426

\bibitem[{{Bahcall} {et~al.}(1993){Bahcall}, {Bergeron}, {Boksenberg},
  {Hartig}, {Jannuzi}, {Kirhakos}, {Sargent}, {Savage}, {Schneider},
  {Turnshek}, {Weymann}, \& {Wolfe}}]{Bah93}
{Bahcall}, J.~N., {Bergeron}, J., {Boksenberg}, A., {Hartig}, G.~F., {Jannuzi},
  B.~T., {Kirhakos}, S., {Sargent}, W.~L.~W., {Savage}, B.~D., {Schneider},
  D.~P., {Turnshek}, D.~A., {Weymann}, R.~J., \& {Wolfe}, A.~M. 1993, \apjs,
  87, 1

\bibitem[{{Bajtlik} {et~al.}(1988){Bajtlik}, {Duncan}, \& {Ostriker}}]{Bat88}
{Bajtlik}, S., {Duncan}, R.~C., \& {Ostriker}, J.~P. 1988, \apj, 327, 570

\bibitem[{{Beers} {et~al.}(1990){Beers}, {Flynn}, \& {Gebhardt}}]{Bee90}
{Beers}, T.~C., {Flynn}, K., \& {Gebhardt}, K. 1990, \aj, 100, 32

\bibitem[{{Bland-Hawthorn} {et~al.}(1997){Bland-Hawthorn}, {Freeman}, \&
  {Quinn}}]{Bla97}
{Bland-Hawthorn}, J., {Freeman}, K.~C., \& {Quinn}, P.~J. 1997, \apj, 490, 143

\bibitem[{{Bochkarev} \& {Sunyaev}(1977)}]{Boc77}
{Bochkarev}, N.~G., \& {Sunyaev}, R.~A. 1977, \azh, 54, 957

\bibitem[{{Bouwens} {et~al.}(2009){Bouwens}, {Illingworth}, {Franx}, {Chary},
  {Meurer}, {Conselice}, {Ford}, {Giavalisco}, \& {van Dokkum}}]{Bou09}
{Bouwens}, R.~J., {Illingworth}, G.~D., {Franx}, M., {Chary}, R., {Meurer},
  G.~R., {Conselice}, C.~J., {Ford}, H., {Giavalisco}, M., \& {van Dokkum}, P.
  2009, \apj, 705, 936

\bibitem[{{Byun} {et~al.}(1994){Byun}, {Freeman}, \& {Kylafis}}]{Byu94}
{Byun}, Y.~I., {Freeman}, K.~C., \& {Kylafis}, N.~D. 1994, \apj, 432, 114

\bibitem[{{Calzetti} {et~al.}(1994){Calzetti}, {Kinney}, \&
  {Storchi-Bergmann}}]{Cal94}
{Calzetti}, D., {Kinney}, A.~L., \& {Storchi-Bergmann}, T. 1994, \apj, 429, 582

\bibitem[{{Carignan} \& {Purton}(1998)}]{Car98}
{Carignan}, C., \& {Purton}, C. 1998, \apj, 506, 125

\bibitem[{{Carswell} {et~al.}(1982){Carswell}, {Whelan}, {Smith}, {Boksenberg},
  \& {Tytler}}]{Car82}
{Carswell}, R.~F., {Whelan}, J.~A.~J., {Smith}, M.~G., {Boksenberg}, A., \&
  {Tytler}, D. 1982, \mnras, 198, 91

\bibitem[{{Cen} {et~al.}(1994){Cen}, {Miralda-Escud{\'e}}, {Ostriker}, \&
  {Rauch}}]{Cen94}
{Cen}, R., {Miralda-Escud{\'e}}, J., {Ostriker}, J.~P., \& {Rauch}, M. 1994,
  \apjl, 437, L9

\bibitem[{{Chen} {et~al.}(2007){Chen}, {Prochaska}, \& {Gnedin}}]{Che07}
{Chen}, H., {Prochaska}, J.~X., \& {Gnedin}, N.~Y. 2007, \apjl, 667, L125

\bibitem[{{Christlein} \& {Zaritsky}(2008)}]{Chr08}
{Christlein}, D., \& {Zaritsky}, D. 2008, \apj, 680, 1053

\bibitem[{{Christlein} {et~al.}(2010){Christlein}, {Zaritsky}, \&
  {Bland-Hawthorn}}]{Chr10}
{Christlein}, D., {Zaritsky}, D., \& {Bland-Hawthorn}, J. 2010, \mnras, 641

\bibitem[{{Corbelli} {et~al.}(1989){Corbelli}, {Schneider}, \&
  {Salpeter}}]{Cor89}
{Corbelli}, E., {Schneider}, S.~E., \& {Salpeter}, E.~E. 1989, \aj, 97, 390

\bibitem[{{Croft} {et~al.}(1998){Croft}, {Weinberg}, {Katz}, \&
  {Hernquist}}]{Cro98}
{Croft}, R.~A.~C., {Weinberg}, D.~H., {Katz}, N., \& {Hernquist}, L. 1998,
  \apj, 495, 44

\bibitem[{{Dav{\'e}} \& {Tripp}(2001)}]{Dav01}
{Dav{\'e}}, R., \& {Tripp}, T.~M. 2001, \apj, 553, 528

\bibitem[{{Dierckx}(1993)}]{Die93}
{Dierckx}, P. 1993, {Curve and surface fitting with splines} (Monographs on
  Numerical Analysis, Oxford: Clarendon, |c1993)

\bibitem[{{Donahue} {et~al.}(1995){Donahue}, {Aldering}, \& {Stocke}}]{Don95}
{Donahue}, M., {Aldering}, G., \& {Stocke}, J.~T. 1995, \apjl, 450, L45+

\bibitem[{{Dove} \& {Shull}(1994)}]{Dov94}
{Dove}, J.~B., \& {Shull}, J.~M. 1994, \apj, 423, 196

\bibitem[{{Efstathiou}(1992)}]{Efs92}
{Efstathiou}, G. 1992, \mnras, 256, 43P

\bibitem[{{Fardal} {et~al.}(1998){Fardal}, {Giroux}, \& {Shull}}]{Far98}
{Fardal}, M.~A., {Giroux}, M.~L., \& {Shull}, J.~M. 1998, \aj, 115, 2206

\bibitem[{{Faucher-Gigu{\`e}re}
  {et~al.}(2008{\natexlab{a}}){Faucher-Gigu{\`e}re}, {Lidz}, {Hernquist}, \&
  {Zaldarriaga}}]{Fau08b}
{Faucher-Gigu{\`e}re}, C., {Lidz}, A., {Hernquist}, L., \& {Zaldarriaga}, M.
  2008{\natexlab{a}}, \apjl, 682, L9

\bibitem[{{Faucher-Gigu{\`e}re}
  {et~al.}(2008{\natexlab{b}}){Faucher-Gigu{\`e}re}, {Lidz}, {Hernquist}, \&
  {Zaldarriaga}}]{Fau08a}
---. 2008{\natexlab{b}}, \apj, 688, 85

\bibitem[{{Faucher-Gigu{\`e}re}
  {et~al.}(2008{\natexlab{c}}){Faucher-Gigu{\`e}re}, {Lidz}, {Zaldarriaga}, \&
  {Hernquist}}]{Fau08c}
{Faucher-Gigu{\`e}re}, C., {Lidz}, A., {Zaldarriaga}, M., \& {Hernquist}, L.
  2008{\natexlab{c}}, \apj, 673, 39

\bibitem[{{Faucher-Gigu{\`e}re} {et~al.}(2009){Faucher-Gigu{\`e}re}, {Lidz},
  {Zaldarriaga}, \& {Hernquist}}]{Fau09}
---. 2009, \apj, 703, 1416

\bibitem[{{Felten} \& {Bergeron}(1969)}]{Fel69}
{Felten}, J.~E., \& {Bergeron}, J. 1969, \aplett, 4, 155

\bibitem[{{Gallego} {et~al.}(1995){Gallego}, {Zamorano}, {Aragon-Salamanca}, \&
  {Rego}}]{Gal95}
{Gallego}, J., {Zamorano}, J., {Aragon-Salamanca}, A., \& {Rego}, M. 1995,
  \apjl, 455, L1+

\bibitem[{{Garc{\'{\i}}a-Ruiz} {et~al.}(2002){Garc{\'{\i}}a-Ruiz}, {Sancisi},
  \& {Kuijken}}]{Gar02}
{Garc{\'{\i}}a-Ruiz}, I., {Sancisi}, R., \& {Kuijken}, K. 2002, \aap, 394, 769

\bibitem[{{Gnedin} {et~al.}(2008){Gnedin}, {Kravtsov}, \& {Chen}}]{Gne08}
{Gnedin}, N.~Y., {Kravtsov}, A.~V., \& {Chen}, H. 2008, \apj, 672, 765

\bibitem[{{Haardt} \& {Madau}(1996)}]{Haa96}
{Haardt}, F., \& {Madau}, P. 1996, \apj, 461, 20

\bibitem[{{Herbert-Fort} {et~al.}(2010){Herbert-Fort}, {Zaritsky},
  {Christlein}, \& {Kannappan}}]{Her10}
{Herbert-Fort}, S., {Zaritsky}, D., {Christlein}, D., \& {Kannappan}, S.~J.
  2010, \apj, 715, 902

\bibitem[{{Hernquist} \& {Springel}(2003)}]{Her03}
{Hernquist}, L., \& {Springel}, V. 2003, \mnras, 341, 1253

\bibitem[{{Hill} {et~al.}(2008)}]{Hil08a}
{Hill}, G.~J., {et~al.} 2008, in \procspie, Vol. 7014, 231

\bibitem[{{Hopkins}(2004)}]{Hop04}
{Hopkins}, A.~M. 2004, \apj, 615, 209

\bibitem[{{Hopkins} {et~al.}(2007){Hopkins}, {Richards}, \&
  {Hernquist}}]{Hop07}
{Hopkins}, P.~F., {Richards}, G.~T., \& {Hernquist}, L. 2007, \apj, 654, 731

\bibitem[{{Huchtmeier}(1989)}]{Huc89}
{Huchtmeier}, R. 1989, {A General Catalog of HI Observations of Galaxies. The
  Reference Catalog.}, ed. {Huchtmeier, W.~K., Richter, O.-G.}

\bibitem[{{Hui} \& {Gnedin}(1997)}]{Hui97}
{Hui}, L., \& {Gnedin}, N.~Y. 1997, \mnras, 292, 27

\bibitem[{{Kelson}(2003)}]{Kel03}
{Kelson}, D.~D. 2003, \pasp, 115, 688

\bibitem[{{Kennicutt}(1989)}]{Ken89}
{Kennicutt}, Jr., R.~C. 1989, \apj, 344, 685

\bibitem[{{Kulkarni} \& {Fall}(1993)}]{Kul93}
{Kulkarni}, V.~P., \& {Fall}, S.~M. 1993, \apjl, 413, L63

\bibitem[{{Lee} {et~al.}(2009){Lee}, {Gil de Paz}, {Tremonti}, {Kennicutt},
  {Salim}, {Bothwell}, {Calzetti}, {Dalcanton}, {Dale}, {Engelbracht}, {Funes},
  {Johnson}, {Sakai}, {Skillman}, {van Zee}, {Walter}, \& {Weisz}}]{Lee09}
{Lee}, J.~C., {Gil de Paz}, A., {Tremonti}, C., {Kennicutt}, R.~C., {Salim},
  S., {Bothwell}, M., {Calzetti}, D., {Dalcanton}, J., {Dale}, D.,
  {Engelbracht}, C., {Funes}, S.~J.~J.~G., {Johnson}, B., {Sakai}, S.,
  {Skillman}, E., {van Zee}, L., {Walter}, F., \& {Weisz}, D. 2009, \apj, 706,
  599

\bibitem[{{Loeb} \& {Eisenstein}(1995)}]{Loe95}
{Loeb}, A., \& {Eisenstein}, D.~J. 1995, \apj, 448, 17

\bibitem[{{Madau} {et~al.}(1999){Madau}, {Haardt}, \& {Rees}}]{Mad99}
{Madau}, P., {Haardt}, F., \& {Rees}, M.~J. 1999, \apj, 514, 648

\bibitem[{{Madsen} {et~al.}(2001){Madsen}, {Reynolds}, {Haffner}, {Tufte}, \&
  {Maloney}}]{Mad01}
{Madsen}, G.~J., {Reynolds}, R.~J., {Haffner}, L.~M., {Tufte}, S.~L., \&
  {Maloney}, P.~R. 2001, \apjl, 560, L135

\bibitem[{{Maloney}(1993)}]{Mal93}
{Maloney}, P. 1993, \apj, 414, 41

\bibitem[{{Massey} {et~al.}(1988){Massey}, {Strobel}, {Barnes}, \&
  {Anderson}}]{Mas88}
{Massey}, P., {Strobel}, K., {Barnes}, J.~V., \& {Anderson}, E. 1988, \apj,
  328, 315

\bibitem[{{Matthews} {et~al.}(1999){Matthews}, {Gallagher}, \& {van
  Driel}}]{Mat99}
{Matthews}, L.~D., {Gallagher}, III, J.~S., \& {van Driel}, W. 1999, \aj, 118,
  2751

\bibitem[{{McCall} {et~al.}(1985){McCall}, {Rybski}, \& {Shields}}]{McC85}
{McCall}, M.~L., {Rybski}, P.~M., \& {Shields}, G.~A. 1985, \apjs, 57, 1

\bibitem[{{O'Donnell}(1994)}]{Odo94}
{O'Donnell}, J.~E. 1994, \apj, 422, 158

\bibitem[{{Oosterloo} {et~al.}(2007){Oosterloo}, {Fraternali}, \&
  {Sancisi}}]{Oos07}
{Oosterloo}, T., {Fraternali}, F., \& {Sancisi}, R. 2007, \aj, 134, 1019

\bibitem[{{Osterbrock} \& {Ferland}(2006)}]{Ost06}
{Osterbrock}, D.~E., \& {Ferland}, G.~J. 2006, {Astrophysics of gaseous nebulae
  and active galactic nuclei}

\bibitem[{{Rauch} {et~al.}(1997{\natexlab{a}}){Rauch}, {Haehnelt}, \&
  {Steinmetz}}]{Rau97a}
{Rauch}, M., {Haehnelt}, M.~G., \& {Steinmetz}, M. 1997{\natexlab{a}}, \apj,
  481, 601

\bibitem[{{Rauch} {et~al.}(1997{\natexlab{b}}){Rauch}, {Miralda-Escude},
  {Sargent}, {Barlow}, {Weinberg}, {Hernquist}, {Katz}, {Cen}, \&
  {Ostriker}}]{Rau97b}
{Rauch}, M., {Miralda-Escude}, J., {Sargent}, W.~L.~W., {Barlow}, T.~A.,
  {Weinberg}, D.~H., {Hernquist}, L., {Katz}, N., {Cen}, R., \& {Ostriker},
  J.~P. 1997{\natexlab{b}}, \apj, 489, 7

\bibitem[{{Schirber} \& {Bullock}(2003)}]{Sch03}
{Schirber}, M., \& {Bullock}, J.~S. 2003, \apj, 584, 110

\bibitem[{{Schlegel} {et~al.}(1998){Schlegel}, {Finkbeiner}, \&
  {Davis}}]{Sch98}
{Schlegel}, D.~J., {Finkbeiner}, D.~P., \& {Davis}, M. 1998, \apj, 500, 525

\bibitem[{{Shapley} {et~al.}(2006){Shapley}, {Steidel}, {Pettini},
  {Adelberger}, \& {Erb}}]{Sha06}
{Shapley}, A.~E., {Steidel}, C.~C., {Pettini}, M., {Adelberger}, K.~L., \&
  {Erb}, D.~K. 2006, \apj, 651, 688

\bibitem[{{Shull} {et~al.}(1999){Shull}, {Roberts}, {Giroux}, {Penton}, \&
  {Fardal}}]{Shu99}
{Shull}, J.~M., {Roberts}, D., {Giroux}, M.~L., {Penton}, S.~V., \& {Fardal},
  M.~A. 1999, \aj, 118, 1450

\bibitem[{{Siana} {et~al.}(2010){Siana}, {Teplitz}, {Ferguson}, {Brown},
  {Giavalisco}, {Dickinson}, {Chary}, {de Mello}, {Conselice}, {Bridge},
  {Gardner}, {Colbert}, \& {Scarlata}}]{Sia10}
{Siana}, B., {Teplitz}, H.~I., {Ferguson}, H.~C., {Brown}, T.~M., {Giavalisco},
  M., {Dickinson}, M., {Chary}, R., {de Mello}, D.~F., {Conselice}, C.~J.,
  {Bridge}, C.~R., {Gardner}, J.~P., {Colbert}, J.~W., \& {Scarlata}, C. 2010,
  ArXiv e-prints

\bibitem[{{Stocke} {et~al.}(1991){Stocke}, {Case}, {Donahue}, {Shull}, \&
  {Snow}}]{Sto91}
{Stocke}, J.~T., {Case}, J., {Donahue}, M., {Shull}, J.~M., \& {Snow}, T.~P.
  1991, \apj, 374, 72

\bibitem[{{Sunyaev}(1969)}]{Sun69}
{Sunyaev}, R.~A. 1969, \aplett, 3, 33

\bibitem[{{Thilker} {et~al.}(2007){Thilker}, {Bianchi}, {Meurer}, {Gil de Paz},
  {Boissier}, {Madore}, {Boselli}, {Ferguson}, {Mu{\~n}oz-Mateos}, {Madsen},
  {Hameed}, {Overzier}, {Forster}, {Friedman}, {Martin}, {Morrissey}, {Neff},
  {Schiminovich}, {Seibert}, {Small}, {Wyder}, {Donas}, {Heckman}, {Lee},
  {Milliard}, {Rich}, {Szalay}, {Welsh}, \& {Yi}}]{Thi07}
{Thilker}, D.~A., {Bianchi}, L., {Meurer}, G., {Gil de Paz}, A., {Boissier},
  S., {Madore}, B.~F., {Boselli}, A., {Ferguson}, A.~M.~N., {Mu{\~n}oz-Mateos},
  J.~C., {Madsen}, G.~J., {Hameed}, S., {Overzier}, R.~A., {Forster}, K.,
  {Friedman}, P.~G., {Martin}, D.~C., {Morrissey}, P., {Neff}, S.~G.,
  {Schiminovich}, D., {Seibert}, M., {Small}, T., {Wyder}, T.~K., {Donas}, J.,
  {Heckman}, T.~M., {Lee}, Y., {Milliard}, B., {Rich}, R.~M., {Szalay}, A.~S.,
  {Welsh}, B.~Y., \& {Yi}, S.~K. 2007, \apjs, 173, 538

\bibitem[{{Tully} {et~al.}(2006){Tully}, {Rizzi}, {Dolphin}, {Karachentsev},
  {Karachentseva}, {Makarov}, {Makarova}, {Sakai}, \& {Shaya}}]{Tul06}
{Tully}, R.~B., {Rizzi}, L., {Dolphin}, A.~E., {Karachentsev}, I.~D.,
  {Karachentseva}, V.~E., {Makarov}, D.~I., {Makarova}, L., {Sakai}, S., \&
  {Shaya}, E.~J. 2006, \aj, 132, 729

\bibitem[{{Uson} \& {Matthews}(2003)}]{Uso03}
{Uson}, J.~M., \& {Matthews}, L.~D. 2003, \aj, 125, 2455

\bibitem[{{van Gorkom}(1993)}]{vGo93}
{van Gorkom}, J. 1993, in Astrophysics and Space Science Library, Vol. 188, The
  Environment and Evolution of Galaxies, ed. {J.~M.~Shull \& H.~A.~Thronson},
  345--+

\bibitem[{{Vogel} {et~al.}(1995){Vogel}, {Weymann}, {Rauch}, \&
  {Hamilton}}]{Vog95}
{Vogel}, S.~N., {Weymann}, R., {Rauch}, M., \& {Hamilton}, T. 1995, \apj, 441,
  162

\bibitem[{{Walsh} {et~al.}(1997){Walsh}, {Staveley-Smith}, \&
  {Oosterloo}}]{Wal97}
{Walsh}, W., {Staveley-Smith}, L., \& {Oosterloo}, T. 1997, \aj, 113, 1591

\bibitem[{{Werk} {et~al.}(2010{\natexlab{a}}){Werk}, {Putman}, {Meurer},
  {Ryan-Weber}, {Kehrig}, {Thilker}, {Bland-Hawthorn}, {Drinkwater},
  {Kennicutt}, {Wong}, {Freeman}, {Oey}, {Dopita}, {Doyle}, {Ferguson},
  {Hanish}, {Heckman}, {Kilborn}, {Kim}, {Knezek}, {Koribalski}, {Meyer},
  {Smith}, \& {Zwaan}}]{Wer10a}
{Werk}, J.~K., {Putman}, M.~E., {Meurer}, G.~R., {Ryan-Weber}, E.~V., {Kehrig},
  C., {Thilker}, D.~A., {Bland-Hawthorn}, J., {Drinkwater}, M.~J., {Kennicutt},
  R.~C., {Wong}, O.~I., {Freeman}, K.~C., {Oey}, M.~S., {Dopita}, M.~A.,
  {Doyle}, M.~T., {Ferguson}, H.~C., {Hanish}, D.~J., {Heckman}, T.~M.,
  {Kilborn}, V.~A., {Kim}, J.~H., {Knezek}, P.~M., {Koribalski}, B., {Meyer},
  M., {Smith}, R.~C., \& {Zwaan}, M.~A. 2010{\natexlab{a}}, \aj, 139, 279

\bibitem[{{Werk} {et~al.}(2010{\natexlab{b}}){Werk}, {Putman}, {Meurer},
  {Thilker}, {Allen}, {Bland-Hawthorn}, {Kravtsov}, \& {Freeman}}]{Wer10b}
{Werk}, J.~K., {Putman}, M.~E., {Meurer}, G.~R., {Thilker}, D.~A., {Allen},
  R.~J., {Bland-Hawthorn}, J., {Kravtsov}, A., \& {Freeman}, K.
  2010{\natexlab{b}}, \apj, 715, 656

\bibitem[{{Weymann} {et~al.}(2001){Weymann}, {Vogel}, {Veilleux}, \&
  {Epps}}]{Wey01}
{Weymann}, R.~J., {Vogel}, S.~N., {Veilleux}, S., \& {Epps}, H.~W. 2001, \apj,
  561, 559

\end{thebibliography}
\bibliographystyle{apj}   

\begin{deluxetable}{clllllllll}
\tabletypesize{\scriptsize}
\tablecaption{HI based model parameters and H$\alpha$ surface brightness predictions\label{tab_MC}\tablenotemark{$\ast$}}
\tablewidth{0pt}
\tablehead{
\colhead{Galaxy} & \colhead{n$_0$} & \colhead{h$_z$} & \colhead{h$_r$} & \colhead{i} & \colhead{PA} & \colhead{$\mu_0$} & \colhead{$(\mu \otimes S)_{0}$} & \colhead{$\bar{\mu}$} & \colhead{$\mu_{\mbox{HI}}$} \\
& \colhead{(cm$^{-3}$)} & \colhead{(pc)} & \colhead{(kpc)} & \colhead{(\arcdeg)} & \colhead{(\arcdeg)} & \colhead{\tablenotemark{$\dagger$}} & \colhead{\tablenotemark{$\ddagger$}} & \colhead{\tablenotemark{$\dagger\dagger$}} & \colhead{\tablenotemark{$\ast\ast$}} \\
}
\startdata
UGC~7321&3.3$^{+3.5}_{-1.7}$&426.$^{+120.}_{-88.}d_{10}$&2.12$^{+0.25}_{-0.16}d_{10}$&82.8$^{+0.9}_{-0.6}$&-100.1$\pm$0.1 &18.4
$^{+1.0}_{-0.9}$&16.7$^{+1.1}_{-0.7}$&16.6$^{+1.0}_{-0.3}$&22.5$^{+4.3}_{-1.8}$\\
UGC~1281&3.8$^{+3.2}_{-2.6}$&303.$^{+70.}_{-58.}d_{5}$&1.17$^{+0.19}_{-0.14}d_{5}$&84.9$^{+4.0}_{-1.3}$&-141.3$\pm$0.3 &21.4
$^{+12.1}_{-2.8}$&19.4$^{+5.6}_{-2.4}$&17.9$^{+1.7}_{-1.1}$&13.4$^{+6.1}_{-2.0}$\\
\enddata
\tablenotetext{$\ast$}{Fit under the restricted radial ranges shown in 
Figure \ref{fig_MC} assuming $\Gamma=4\times 10^{-14}$ s$^{-1}$ and $\beta=1.8$.}
\tablenotetext{$\dagger$}{10$^{-20}$ erg/s/cm$^2$/\sq\arcsec}
\tablenotetext{$\ddagger$}{10$^{-20}$ erg/s/cm$^2$/\sq\arcsec, smoothed by a circular 10\arcsec\ FWHM kernel}
\tablenotetext{$\dagger\dagger$}{10$^{-20}$ erg/s/cm$^2$/\sq\arcsec, average for all fiber positions with 
predicted values of $\mu >$10$^{-19}$ erg/s/cm$^2$/\sq\arcsec}
\tablenotetext{$\ast\ast$}{10$^{-20}$ erg/s/cm$^2$/\sq\arcsec, based on the HI bounded area (Equations 6-8 of 
\citet{Wey01})}
\end{deluxetable}

\begin{deluxetable}{lccccccc}
\tabletypesize{\scriptsize}
\tablecaption{Error budget and limits to the UVB strength\label{tab_lims}}
\tablewidth{0pt}
\tablehead{
\colhead{Co-addition} & \colhead{Poisson} & \colhead{Resolution} & \colhead{Flux} & \colhead{SB} & \colhead{Model} & \colhead{$\Gamma(z=0)$} & \colhead{$\chi^2/\nu$} \\
\colhead{type} & \colhead{error} & \colhead{systematic} & \colhead{calibration} & \colhead{upper limit} & \colhead{systematic} & \colhead{upper limit} & \\
\colhead{(1)} & \colhead{(2)} & \colhead{(3)} & \colhead{(4,\%)} & \colhead{(5)} & \colhead{(6,\%)} & \colhead{(7)} & \colhead{(8)} \\
}
\startdata
UGC~7321 each single fiber & 11.0 & 0.8 & 8.9 & 64 & +5.4/-4.9 & 15 & 353/468\\
UGC~7321 smoothed & 2.8 & 0.4 & 8.9 & 17 & +6.6/-4.2 & 4.4 & 257/454\\
UGC~7321 radio bound stack & 1.8 & 0.4 & 8.9 & 12 & +19.3/-7.9 & 2.3 & 545/462\\
UGC~7321 full stack & 0.9 & 0.4 & 8.9 & 7.1 & +6.0/-1.8 & 1.7 & 497/454\\
UGC~1281 each single fiber & 18.6 & 29.5 & 4.3 & 250 & +57/-13 & 53 & 68/136\\
UGC~1281 smoothed & 6.6 & 8.9 & 4.3 & 81 & +29/-12 & 19 & 218/136\\
UGC~1281 radio bound stack & 6.0 & 8.9 & 4.3 & 78 & +46/-15 & 27 & 261/134\\
UGC~1281 full stack & 2.0 & 8.9 & 4.3 & 57 & +9.5/-6.1 & 14 & 50/134\\
\enddata
\tablenotetext{(1)}{Detection and model method. Smoothed refers to a 10\arcsec\ FWHM Gaussian, circular kernel.}
\tablenotetext{(2)}{1$\sigma$ (10$^{-20}$ erg/s/cm$^2$/\sq\arcsec) in the spectral data from Poisson noise.}
\tablenotetext{(3)}{1$\sigma$ (10$^{-20}$ erg/s/cm$^2$/\sq\arcsec) in the spectral data from spectral resolution or sky line variation. See \S \ref{sec_err}.}
\tablenotetext{(4)}{1$\sigma$ (\%) flux calibration systematic including 2.1\% for airmass/extinction error.}
\tablenotetext{(5)}{5$\sigma$ (10$^{-20}$ erg/s/cm$^2$/\sq\arcsec) limit in surface brightness. No detections of 
significance were found, so this limit results simply from summing columns 2 and 3, 
multiply by one plus the percentage in column 4, and finally multiplying by five.}
\tablenotetext{(6)}{1$\sigma$ (\%) model surface brightness systematic. These values 
are derived from the Monte Carlo tests of \S \ref{sec_mod_fit}.}
\tablenotetext{(7)}{5$\sigma$ (10$^{-14}$ s$^{-1}$) total limit assuming $\beta$~=~1.8. The achieved 
H$\alpha$ surface brightness limit is compared to the low bound of the 
modeled H$\alpha$ surface brightness to create this final, linearized estimate from the 
modeled value of $\Gamma=4\times 10^{-14}$ s$^{-1}$. This limit results simply by multiplying column 5 
by one plus the percentage in column 6, 
multiplying by the baseline $\Gamma=4\times 10^{-14}$ s$^{-1}$ value, and dividing by the lower 
bound to either column 7, 8, 9, or 10 in Table \ref{tab_MC} depending on the limit type.}
\tablenotetext{(8)}{$\chi^2$ of all pixels between 6300-6600\AA\ in the spectrum and the degrees of freedom.}
\end{deluxetable}

\end{document}